\documentclass[longauth]{aa}%

\usepackage{graphicx}
\usepackage{txfonts}
\usepackage{siunitx}
\usepackage[switch]{lineno}
\usepackage{subfig}
\usepackage{overpic}
\usepackage{xspace}
\usepackage{float}
\usepackage{booktabs}
\usepackage[allcolors=blue]{hyperref}
\usepackage{orcidlink}
\hypersetup{colorlinks=true,linkcolor=blue,urlcolor=blue}

\newcommand{\msun}{\hbox{M$_{\odot}$}\xspace}
\newcommand{\herschel}{\textit{Herschel}\xspace}

\newcommand{\kms}{\,\rm km\,s$^{-1}$\xspace}
\newcommand{\kmspc}{\,\rm km\,s$^{-1}$\,pc$^{-1}$\xspace}
\newcommand{\pc}{{\rm pc}}

\newcommand{\vlsr}{V$_{\rm LSR}$\xspace}
\newcommand{\mum}{$\mu$m\xspace}
\newcommand{\degree}{\mbox{$^{\circ}$}\xspace}

\newcommand{\nhp}{N$_2$H$^+$\xspace}%N2Hp
\newcommand{\co}{C$^{18}$O\xspace} %C18O
\newcommand{\dv}{$\Delta$V\xspace}
\newcommand{\mrate}{$\dot{M}_{{\rm in}}$\xspace}

\newcommand{\hh}{\ion{H}{ii}\xspace}
\newcommand{\col}[1]{\multicolumn{1}{c}{#1}}
\newcommand{\novel}{\multicolumn{1}{c}{\hspace{1.015cm}\textemdash}}
\newcommand{\notype}{\multicolumn{1}{l}{\hspace{0.775cm}\textemdash}}
\newcommand{\notypecom}{\multicolumn{1}{l}{\hspace{1cm}\textemdash}}

\newcolumntype{N}{S[table-format=2.2]}

\DeclareFontFamily{U}  {MnSymbolC}{}
\DeclareFontShape{U}{MnSymbolC}{m}{n}{
    <-6>  MnSymbolC5
   <6-7>  MnSymbolC6
   <7-8>  MnSymbolC7
   <8-9>  MnSymbolC8
   <9-10> MnSymbolC9
  <10-12> MnSymbolC10
  <12->   MnSymbolC12}{}
\DeclareFontShape{U}{MnSymbolC}{b}{n}{
    <-6>  MnSymbolC-Bold5
   <6-7>  MnSymbolC-Bold6
   <7-8>  MnSymbolC-Bold7
   <8-9>  MnSymbolC-Bold8
   <9-10> MnSymbolC-Bold9
  <10-12> MnSymbolC-Bold10
  <12->   MnSymbolC-Bold12}{}
\DeclareSymbolFont{MnSymbolC}         {U}  {MnSymbolC}{m}{n}
\SetSymbolFont{MnSymbolC}       {bold}{U}  {MnSymbolC}{b}{n}
\DeclareMathSymbol{\downY}{\mathrel}{MnSymbolC}{41}

\DeclareFontFamily{U}  {MnSymbolF}{}
\DeclareFontShape{U}{MnSymbolF}{m}{n}{
    <-6>  MnSymbolF5
   <6-7>  MnSymbolF6
   <7-8>  MnSymbolF7
   <8-9>  MnSymbolF8
   <9-10> MnSymbolF9
  <10-12> MnSymbolF10
  <12->   MnSymbolF12}{}
\DeclareFontShape{U}{MnSymbolF}{b}{n}{
    <-6>  MnSymbolF-Bold5
   <6-7>  MnSymbolF-Bold6
   <7-8>  MnSymbolF-Bold7
   <8-9>  MnSymbolF-Bold8
   <9-10> MnSymbolF-Bold9
  <10-12> MnSymbolF-Bold10
  <12->   MnSymbolF-Bold12}{}
\DeclareSymbolFont{MnSymbolF}         {U}  {MnSymbolF}{m}{n}
\SetSymbolFont{MnSymbolF}       {bold}{U}  {MnSymbolF}{b}{n}
\DeclareMathSymbol{\xmarker}{\mathrel}{MnSymbolF}{2}

\begin{document}
   \title{ALMA-IMF XIII: \nhp kinematic analysis on the intermediate
   protocluster G353.41}
   \author{R.\ H.\ Álvarez-Gutiérrez \inst{1}\orcidlink{0000-0002-9386-8612},
          A.\ M. Stutz \inst{1}\orcidlink{0000-0003-2300-8200},
          N.\ Sandoval-Garrido \inst{1}\orcidlink{0000-0001-9600-2796},
          F.\ Louvet \inst{2}\orcidlink{0000-0003-3814-4424},
          F.\ Motte  \inst{2}\orcidlink{0000-0003-1649-8002},
          R.\ Galván-Madrid \inst{3}\orcidlink{0000-0003-1480-4643},
          N.\ Cunningham \inst{2,4}\orcidlink{0000-0003-3152-8564},
          P.\ Sanhueza\inst{5,6}\orcidlink{0000-0002-7125-7685},
          M.\ Bonfand \inst{7}\orcidlink{0000-0001-6551-6444},
          S.\ Bontemps \inst{8}\orcidlink{0000-0002-4093-7178},
          A.\ Gusdorf \inst{9,10}\orcidlink{0000-0002-0354-1684},
          A.\ Ginsburg \inst{11}\orcidlink{0000-0001-6431-9633},
          T.\ Csengeri \inst{8}\orcidlink{0000-0002-6018-1371},
          S.\ D.\ Reyes \inst{1,12}\orcidlink{0000-0003-0276-5368},
          J.\ Salinas \inst{1}\orcidlink{0009-0009-4976-4320},
          T.\ Baug \inst{13}\orcidlink{0000-0003-0295-6586},
          L.\ Bronfman \inst{14}\orcidlink{0000-0002-9574-8454},
          G.\ Busquet \inst{15,16,17}\orcidlink{0000-0002-2189-6278},
          D.\ J. Díaz-González \inst{3}\orcidlink{0000-0002-6325-8717},
          M.\ Fernandez-Lopez \inst{18}\orcidlink{0000-0001-5811-0454},
          A.\ Guzmán \inst{19} \orcidlink{0000-0003-0990-8990},
          A.\ Koley \inst{1}\orcidlink{0000-0003-2713-0211},
          H.-L.\ Liu \inst{20}\orcidlink{0000-0003-3343-9645},
          F.\ A.\ Olguin \inst{21}\orcidlink{0000-0002-8250-6827},
          M.\ Valeille-Manet \inst{8}\orcidlink{0009-0005-5343-1888},
          F.\ Wyrowski \inst{22}\orcidlink{0000-0003-4516-3981}
          }
   \institute{Departamento de Astronom\'{i}a, Universidad de Concepci\'{o}n,
         Casilla 160-C, Concepci\'{o}n, Chile %yo,amy,nico,javi,atanu
         \email{rodralvarezz@gmail.com}
         \and %fabien,fred,nichol
         Univ. Grenoble Alpes, CNRS, IPAG, 38000 Grenoble, France 
         \and
         Instituto de Radioastronomía y Astrofísica, Universidad Nacional Autónoma de México, Morelia, Michoacán 58089, México
         \and
         SKA Observatory, %nichol
         Jodrell Bank, Lower Withington, Macclesfield, SK11 9FT, United Kingdom
         \and
         National Astronomical Observatory of Japan, 2-21-1 Osawa, Mitaka, Tokyo 181-8588, Japan%patricio
         \and
         %patricio
         Astronomical Science Program, The Graduate University for Advanced Studies, SOKENDAI, 2-21-1 Osawa, Mitaka, Tokyo 181-8588, Japan
         \and
         Departments of Astronomy, University of Virginia, Charlottesville, VA 22904, USA%melisse
         \and%maxime %timea         
         Laboratoire d'astrophysique de Bordeaux, Univ. Bordeaux, CNRS, B18N, allée Geoffroy Saint-Hilaire, F-33615 Pessac, France
         \and%antoine
         Laboratoire de Physique de l'\'{E}cole Normale Sup\'{e}rieure, ENS, 
         Universit\'{e} PSL, CNRS, Sorbonne Universit\'{e}, Universit\'{e} Paris 
         Cit\'{e}, F-75005, Paris, France
         \and%antoine
         Observatoire de Paris, PSL University, Sorbonne Universit\'{e}, 
         LERMA, 75014, Paris, France
         \and%adam
         Department of Astronomy, University of Florida, P.O. Box 112055, Gainesville, FL 32611, USA
         \and
         Max Planck Institute for Astronomy, Königstuhl 17, 69117 Heidelberg, Germany %simon
         \and%tapas
         S. N. Bose National Centre for Basic Sciences, Sector-III, Salt Lake, Kolkata 700106, India
         \and%leonardo bronfman
         Astronomy Department, Universidad de Chile, Camino El Observatorio 1515, Las Condes, Santiago, Chile
         \and%gemma
         Departament de Física Quàntica i Astrofísica (FQA), Universitat de Barcelona (UB), Martí i Franquès 1, 08028 Barcelona, Catalonia, Spain
         \and%gemma
         Institut de Ciències del Cosmos (ICCUB), Universitat de Barcelona, Martí i Franquès, 1, 08028, Barcelona, Catalonia, Spain
         \and%gemma
         Institut d'Estudis Espacials de Catalunya (IEEC), Gran Capità, 2-4, 08034 Barcelona, Catalonia, Spain
         \and %manuel
         Instituto Argentino de Radioastronomía (CCT-La Plata, CONICET; CICPBA), C.C. No. 5, 1894, Villa Elisa, Buenos Aires, Argentina         
         \and%andres
         Joint Alma Observatory (JAO), Alonso de Córdova 3107, Vitacura, Santiago, Chile         
         \and%hongli
         School of Physics and Astronomy, Yunnan University, Kunming, 650091, People’s Republic of China
         \and%fernando olguin
         Institute of Astronomy and Department of Physics, National Tsing Hua University, Hsinchu 30013, Taiwan 
         \and%Friedrich
         Max-Planck-Institut für Radioastronomie, Auf dem Hügel 69,
         53121 Bonn, Germany
             }

   \authorrunning{R. H. Álvarez-Gutiérrez et al.}

   \date{Received 11 April 2024 / Accepted 12 June 2024}
   
   \abstract{
The ALMA-IMF Large Program provides multi-tracer observations of
15 Galactic massive protoclusters at matched sensitivity and spatial 
resolution. We focus on the dense gas kinematics of the
G353.41 protocluster traced by N$_2$H$^+$ (1$-$0), with an spatial
resolution $\sim$\,0.02~pc. G353.41, at a distance of $\sim$~2\,kpc, is embedded in a larger scale
($\sim\,8\,$pc) filament and has a mass of 
\text{$\sim2.5\,\times\,10^3$\,M$_{\odot}$} within $1.3\times1.3$~pc$^2$.
We extract the N$_2$H$^+$ (1$-$0) isolated line component and we decompose it 
by fitting up to 3 Gaussian velocity components. This allows us to identify 
velocity structures that are either muddled or impossible to identify in the 
traditional position-velocity diagram.  We identify multiple velocity 
gradients on large ($\sim$~1\,pc) and small scales ($\sim$0.2\,pc).
We find good agreement between the N$_2$H$^+$ velocities and 
the previously reported DCN core velocities, suggesting that cores are
kinematically coupled to the dense gas in which they form.  We measure
9 converging ``V-shaped'' velocity gradients (VGs)
($\sim20\,$km\,s$^{-1}$\,pc$^{-1}$) that are well-resolved 
(sizes $\sim\,0.1\,$pc), located in filaments, which are sometimes 
associated with cores near their point of convergence.
We interpret these V-shapes as inflowing gas feeding the regions near cores 
(the immediate sites of star formation). We estimate 
the timescales associated with V-shapes as VG$^{-1}$, and we interpret them as
inflow timescales.
The average inflow timescale is
$\sim\,67$~kyr, or about twice the free-fall time of cores in the same area 
($\sim\,33\,$kyr) but substantially shorter than protostar lifetime 
estimates ($\sim\,$0.5~Myr). We derive mass accretion 
rates in the range of $(0.35-8.77)\,\times\,10^{-4}$~M$_{\odot}$~yr$^{-1}$.
This feeding might lead to further filament collapse and formation of
new cores.  We suggest that the protocluster is
collapsing on large scales, but the velocity signature of collapse 
is slow compared to pure free-fall. Thus these data 
are consistent with a comparatively slow global protocluster 
contraction under gravity, and faster core formation within, 
suggesting the formation of multiple generations 
of stars over the protocluster lifetime.
   }
      

   \maketitle
%
%________________________________________________________________

\section{Introduction}
\label{sect:intro}

While star clusters have been studied extensively over many decades at
comparatively short wavelengths, their precursors, protoclusters have
not been studied in depth until recently.  Protoclusters (or embedded
clusters) are the gas-dominated maternal environments where star
clusters are born and whose stellar constituents will ultimately
populate the field of our Galaxy. Protoclusters are distinct entities
from star clusters. Both are defined as relatively compact
configurations where the gravity is strong enough to influence the
dynamics of their constituents. But in the latter, there is little to
no gas, and the gravity of the cluster is dominated by the stars themselves.
In protoclusters, in contrast, gravity is dominated by the cold gas in
which the stars themselves are forming  \citep[][]{stutz16,timea17,stutz18,motte2018}. 
Protoclusters are more accessible now than ever before thanks to ALMA and its 
exquisitely high resolution interferometric mm-wave data tracing the cold gas 
where the stars form \citep[][]{sanhueza19,atoms,motte2022}. Inside 
protoclusters we witness the ongoing conversion of gas into compact and 
extremely dense stars, a process mediated by gas filaments 
\citep[][]{stutz18,gonzalez19,alvarez21} feeding gas structures called 
``cores''  \citep[][]{andre10,stutz15,kuznetsova15,kuznetsova18}.
Cores are compact gas mass concentrations, often defined to be of a
size matching the resolution limit of the observations.  In this case, 
we define cores to be order $\sim\,2$~kau, for reasons described below. 

\begin{figure}
\centering
\includegraphics[width=\columnwidth]{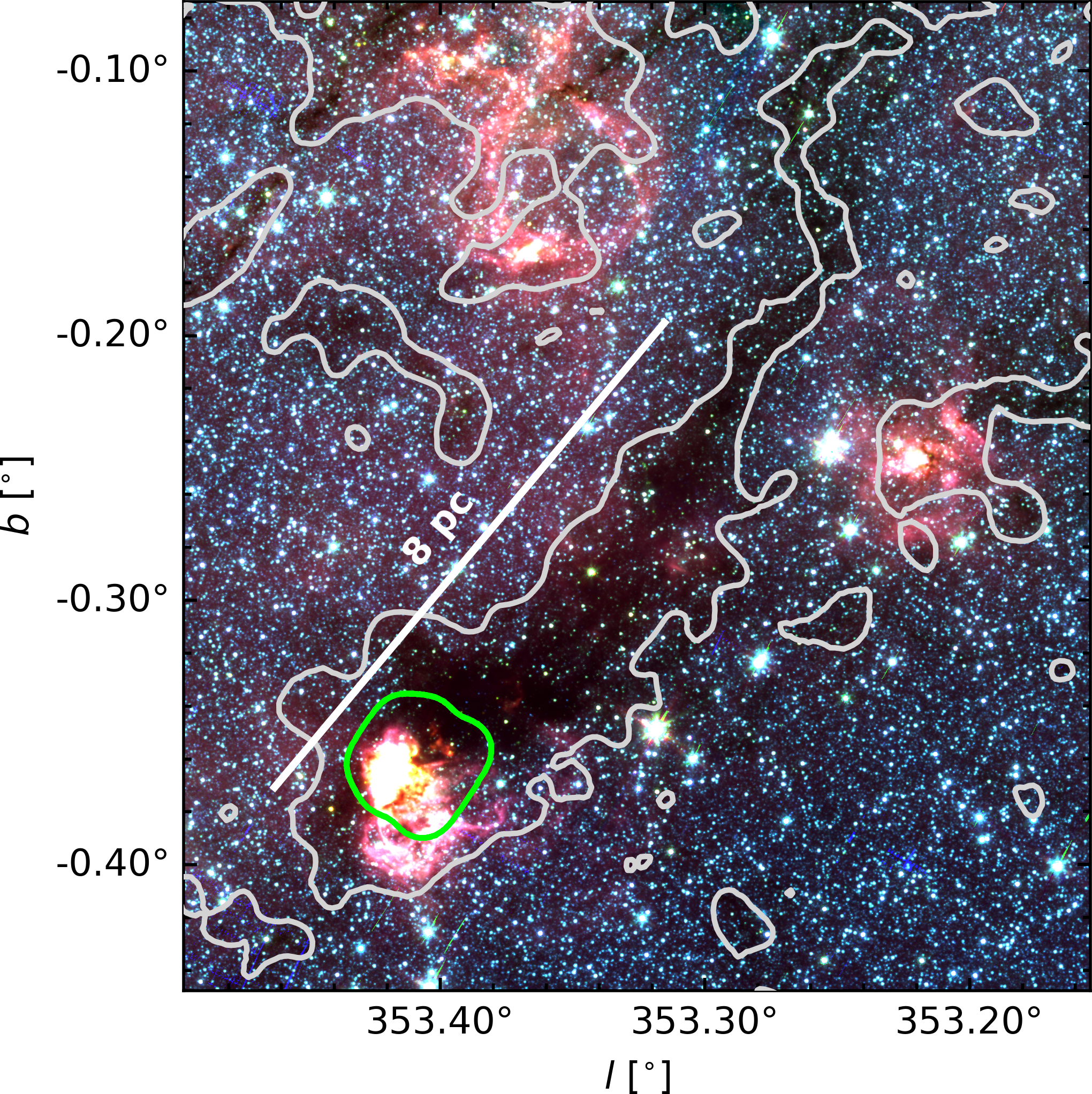}
\caption{Composite image of G353: IRAC 3.6~\mum (in blue),
4.5~\mum (green), and 5.8~\mum (red).
We indicate the ALMA-IMF \nhp (1$-$0) coverage with a light green contour.
We highlight ATLASGAL emission (870\,\mum) at 40~mJy~beam$^{-1}$ with the 
gray contour, corresponding roughly to a \herschel derived $N(H)$ of 
$\sim~5.5\,10^{22}$~cm$^{-2}$.}
\label{fig:rgb}
\end{figure}

In this paper, we focus on the G353.41 protocluster (see Fig.~\ref{fig:rgb}),
and in particular, on the dense gas kinematics observationally 
accessible from the protocluster scale (2.9\,pc$^2$) to the core scale.
We trace this dense and cold gas using the \nhp (1-0) line observed
with ALMA.
Given \nhp is detected at high column densities
\citep[N(H$_2$)$~\gtrsim~10^{22}$\,cm$^{-2}$;][]{tafalla2021}, we gain
access to the inner dense gas "skeleton" of the
protocluster structure, free from confusion induced by lower density
gas. Meanwhile, ALMA permits us to obtain the resolution needed to
trace structures down to the core scales where individual or small
numbers of stars may be forming.

The ALMA-IMF Large Program\footnote{Proposal ID 2017.1.01355.L,
PIs: Motte, Ginsburg, Louvet, Sanhueza} (LP)
maps 15 dense, nearby ($2~-~5.5$~kpc), and massive 
($2~-~32~\times$~10$^3$~\msun) Milky Way
protoclusters down to $\sim$2~kau scales \citep[][]{motte2022},
at matched spatial resolution.
ALMA-IMF provides a large protocluster sample in order to test the
universality of the stellar initial mass function (IMF)
\citep[][]{bastian10,offner14}.
The ALMA-IMF LP also provides a vast catalogue
of molecular lines, in bands 3 ($2.6-3.6$~mm) and 6 ($1.1-1.4$~mm). 
This rich molecular treasure trove allows for a detailed kinematical 
characterization of the gas, protostellar cores, and young stellar objects 
(YSOs) present in these protoclusters.
The current publicly available ALMA-IMF data include, but are not
limited to, continuum maps \citep[][]{ginsburg22b,diaz23}, 12~m data
cubes of all spectral windows \citep[][]{cunningham2023}, core
catalogues \citep[][Louvet et al. submitted]{pouteau2022}, and 
hot core and outflow catalogues \citep[][Valeille-Manet et al. in prep]{cunningham2023,tomas23,towner24,armante24,bonfand24}.
The data products derived from the ALMA-IMF LP allow us to
constrain the different star forming environments,
where we can analyze column densities, temperatures, outflow masses, core
properties, and multi-tracer gas kinematics. This approach 
offers a thorough characterization of the processes taking place 
in these regions.

\begin{table*}
\caption{Relevant parameters of the \nhp 7m+12m imaging}
\label{table:imaging_params}
\centering
\begin{tabular}{c c c c c c c c c}
\hline\hline
Field size & Pixel scale & Beam size & BPA  & $^{a}$RMS & Channel width & RMS velocity range & $^{b}$\vlsr \ \\%& $^{c}$Pbmask\\
%vlsr taken from the reduction script imaging_parameters. %$^{b}$RMS
          &              &           & [°]\ \ & \ [K]   &   [\kms]       &   [\kms]           & [\kms]                \\% & \ [K]
\hline
         176\arcsec$\times$172\arcsec   &      0.72\arcsec & 1.96\arcsec$\times$2.29\arcsec &  80.19  & 0.37     &      0.23 &  [$-$43 ; $-$32],   [0 ; +7]   &      $-$17        \\%& 0.18  \\% 0.68  &
         1.72\,\pc$\times$1.67\,\pc     &      1.44\,kau    & $\sim$4\,kau$\times$4.6\,kau & & & &\\% &\\
\hline
\end{tabular}
\tablefoot{$^{a}$ RMS value at the peak of the RMS distribution. $^{b}$
Obtained from \citet[][]{motte2022}. }
% Table adapted from Stutz et al. (in prep.).
%$^{c}$ Pbmask value used in the imaging of the 7m+12m dataset.}
\end{table*}

\citet[][]{motte2022} present a method of classifying these 15
protoclusters based on their evolutionary stage, 
assuming that they exhibit more \hh regions as they evolve.
They take into account the flux ratio between the 1~mm to 3~mm continuum maps
(S$^{\rm{cloud}}_{\rm{1.3~mm}}/\rm{S}^{\rm{cloud}}_{\rm{3~mm}}$), and the
free-free emission at the frequency of H41$\alpha$
($\sum_{\rm{H41}\alpha}^{\rm{free-free}}$).
They find that as protoclusters evolve,
S$^{\rm{cloud}}_{\rm{1.3~mm}}/\rm{S}^{\rm{cloud}}_{\rm{3~mm}}$ decreases,
while $\sum_{\rm{H41}\alpha}^{\rm{free-free}}$ increases
\citep[][see their Fig.~3]{motte2022}.
Using these constraints, they group their 15 protocluster
as being in a young, intermediate, or evolved evolutionary state.
Out of these 15 regions, we analyze the G353.41 protocluster 
(hereafter G353). 
In Fig.~\ref{fig:rgb} we indicate the ALMA-IMF \nhp (1$-$0) coverage of 
G353 (centered at \text{$\alpha$,$\delta$ (J2000)~=~17:30:26.28,$-$34:41:49.7}) 
and its parent filament
\citep[dark lane traced by ATLASGAL 870\,\mum emission;][]{schuller09}
with light green and gray contours respectively. 
\citet[][]{motte2022} classify this protocluster as being at an 
intermediate evolutionary state, located at $\sim$2~kpc, and hosting a total 
mass of 2.5$\,\times\,10^{3}$~\msun. 
They describe G353 as isolated, without obvious interaction with 
massive nearby stellar clusters. Using moment maps
derived from the \nhp (1$-$0) 12~m dataset they suggest 
the presence of multiple velocity components indicating a 
complex velocity field. They propose that G353 is composed of 
filaments interacting at the central hub.
As presented in \citet[][]{bonfand24}, this region is an 
outlier in the ALMA-IMF hot core sample. Only one weak, 
low-mass ($<$~2~\msun) compact methyl formate source is detected
and it lacks strong emission from complex organic molecules. 
They state that this protocluster is in a chemically poor stage, 
where further characterization of this region is required.

The \nhp (1$-$0) transition ($\nu$~=~93.173809~GHz), given its high critical
density, $n_{crit}\,=\,2\times10^5~\rm{cm}^{-3}$ 
\citep[][]{ungerechts97}, allows us to access the dense gas kinematics 
present in the innermost parts of star forming
regions \citep[][]{caselli2002,bergin2002, tafalla2004, lippok13,storm2014,hacar18,cheng19,gonzalez19,alvarez21}.
The J~=~1$\rightarrow$0 transition presents seven hyperfine components
\citep[][]{cazzoli85,caselli95,caselli2002}. The kinematic analysis of this 
complex emission can be simplified by considering only the well 
separated isolated component 
\citep[93.17631~GHz; $F_1,F~=0,1~\rightarrow~1,2$;][]{cazzoli85}.
Such simplification is convenient to study the
complex velocity fields found at the center of filaments. 
These regions present the densest environments for star formation,
usually presenting multiple, blended velocity components, where the 
velocity distributions exhibit twists, turns, spirals, and wave-like patterns 
\citep[][]{timea11,fernandez14,stutz16,liu19,gonzalez19,henshaw20, alvarez21,sanhueza21, redaelli22,olguin23}.
Recent techniques, such as the intensity-weighted position-velocity 
(PV) diagrams \citep[][]{gonzalez19,alvarez21}, allow us to
characterize processes such as infall, outflow, or rotation present
in these environments, where high spatial and spectral resolution studies
open a window into the small scale gas kinematics of star forming 
regions. 
In addition to the PV diagrams, we can create Position-Position-Velocity 
(PPV) diagrams, in order to identify coherent structures that 
might be both spatially and kinematically associated 
\citep[][]{cheng19,henshaw19,henshaw20,sanhueza21,redaelli22}.

In this paper we investigate the \nhp dense gas kinematics of G353 from large
(protocluster) to small (cores) scales.
In \S~\ref{sect:data} we present the data. In \S~\ref{sect:iso_extraction}
we introduce our \nhp isolated extraction procedure. In
\S~\ref{sect:line_modeling} we model and decompose the multiple velocity
components found in the \nhp isolated component spectra. In \S~\ref{sect:analysis} we show our gas kinematic
analysis, from protocluster to core scales.
In \S~\ref{sect:collapse} we show that G353 might be
under gravitational collapse at small and large scales.
In \S~\ref{sect:mrates} we estimate mass accretion rates for
multiple velocity gradients characterized in our \nhp data.
We discuss our results in \S~\ref{sect:discuss}, and we present
our summary and conclusions in \S~\ref{sect:conclusions}.

\section{Data}
\label{sect:data}

\subsection{ALMA-IMF data}
\label{sect:alma-imf-data}
We make use of the \nhp (1$-$0) 12~m, 7~m, and Total Power 
observations described in \citet[][]{motte2022} for our analysis, 
providing robust uv plane coverage.
We image the combination of the \nhp 7\,m and 12\,m
(from now on called ``7m+12m'') measurement set of G353, using the
publicly available imaging scripts from the ALMA-IMF github
repository\footnote{\href{https://github.com/ALMA-IMF/reduction}
{\url{https://github.com/ALMA-IMF/reduction}}}. These data
are corrected by the primary beam response pattern.
Due to the missing large-scale emission, we find that near the \vlsr
of the protocluster \citep[$-$17\,\kms;][]{wienen2015,motte2022} some 
subregions in the 7m+12m cube present deep negative artifacts
(``negative bowls''). % \citep[][]{pouteau2022}.
To cover all possible uv scales, we combine the \nhp 7m+12m 
continuum-subtracted cube with the Total Power observations from the 
ALMA-IMF LP. We use the \texttt{feather}\footnote{\href{https://casa.nrao.edu/docs/
taskref/feather-task.html}{\url{https://casa.nrao.edu/docs/taskref/feather-task.html}}} task from \texttt{CASA 5.6.0}. 
With this combination, we were able to recover the
missing flux, seen as negative bowls, present in the interferometric-only
data. We produce a fully combined, multi-scale, feathered dataset which
we use for our dense gas kinematic analysis.

\begin{figure}
\centering
\includegraphics[width=\columnwidth]{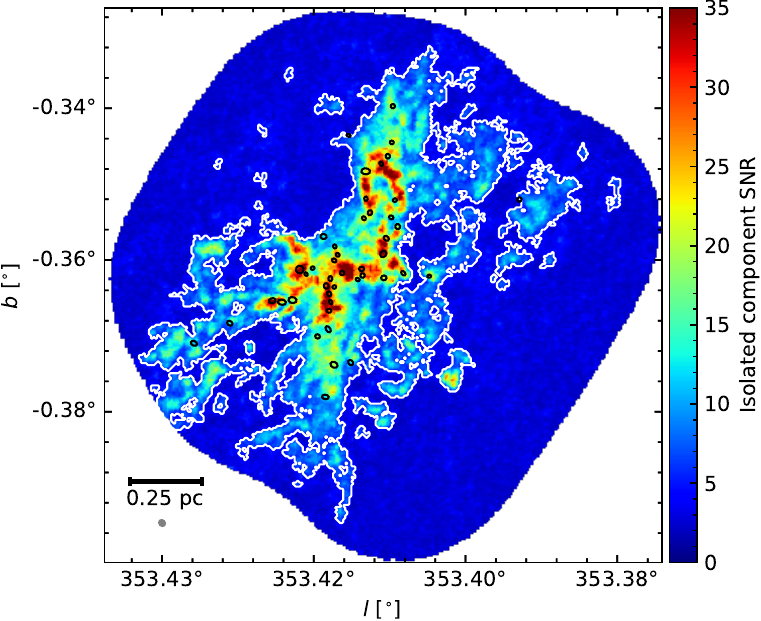}
\caption{G353 \nhp isolated component SNR map. The white contour
indicates the location of data with an isolated component 
SNR~$\geq$~5. We show the location of the 1.3~mm cores presented in 
Louvet et al. (submitted) with black ellipses and are located in regions 
with SNR$~\geq15$. We indicate the beam 
size of these data with a gray ellipse at the bottom left corner.
Outside the SNR contour we make a rough extraction of the isolated 
component (see text). For data inside the SNR contours, we implement a 
procedure based on detection of peaks and valleys, to individually 
extract high ($\geq~5$) SNR isolated components (see 
\S~\ref{sect:iso_extraction}).
}
   \label{fig:snr}
\end{figure}

To estimate and subtract the continuum emission present in the 7m+12m cube,
we use the 
\texttt{imcontsub}\footnote{\href{https://casa.nrao.edu/docs/taskref/
imcontsub-task.html}{\url{https://casa.nrao.edu/docs/taskref/imcontsub-task.html}}}
CASA task. We select the emission-free channels between $-$43\kms and 
$-$33\kms, and set the polynomial degree of the continuum fit 
(\texttt{fitorder}) to 0.
We list relevant final image parameters in Table~\ref{table:imaging_params},
such as the field size, pixel scale, beam size, root-mean-square noise (RMS),
and channel width.

We use 12~m datacubes from \citet[][]{cunningham2023} to compare 
the shock tracers SiO ($5-4$) and CO ($2-1$) to our \nhp kinematic 
analysis. We use DCN and \nhp data to determine core velocities 
(\S~\ref{sect:nhp_vels}). 
To determine total masses in specific regions we use the $N$(H$_2$) 
map from \citet[][]{diaz23}.

\subsection{Core properties from published catalogues}
\label{sect:data-catalogues}
% {\bf explain briefly the use of each dataset}
We use the cores 
catalogue\footnote{Available at \href{almaimf.com}{\url{www.almaimf.com}}} 
from  Louvet et al. (submitted). 
These cores where identified using the \texttt{getsf} algorithm, specialized 
in source extraction on regions with complex filamentary structures 
\citep[][]{mensch21}. This procedure was done using the 1.3~mm continuum maps, 
smoothed at a common resolution of $\sim$2700\,au, obtaining a total of 45 
sources for G353.
We also use the DCN core velocities \citep[15 sources,][]{cunningham2023} and 
the SiO outflow catalogue \citep[16 sources,][]{towner24}
in order to look for correlation between the \nhp gas kinematics and 
cores/outflows position and properties.
It is worth mentioning that, within a radius of $0.3~\pc$ from the
center of G353 \citep[][]{motte2022}, we find 60\% of the 1.3~mm cores 
(27~sources), and $\sim~70\%$ of the cores with DCN velocities and 
SiO outflows (11~sources from each catalogue). Of these 11 outflows, 
7 are ``red'' 3 are ``blue'' (monopolar), and 1 is ``bipolar'' 
\citep[][]{towner24}. 
The presence of these sources might imply a complex velocity field 
in this region, given that cores and outflows disturb the kinematics of the
surrounding gas.

\begin{figure*}
\centering
\includegraphics[width=\textwidth]{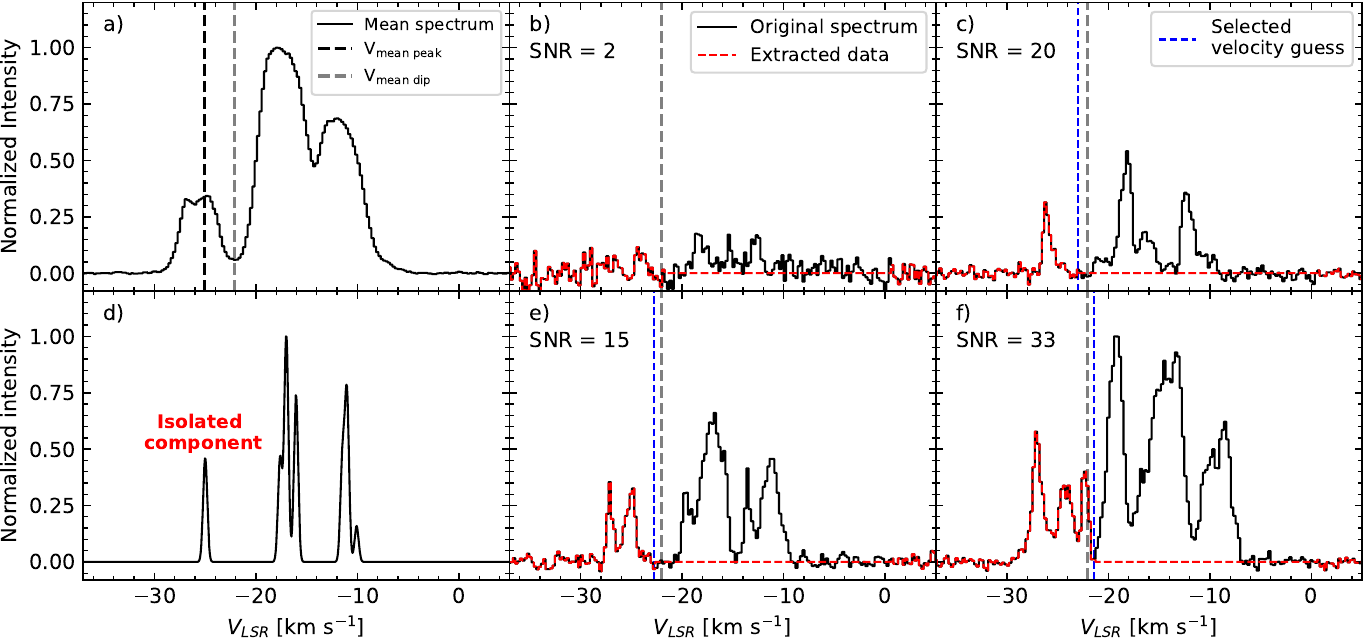}
   \caption{
   {\bf Panel a):} Normalized average \nhp spectrum (solid black line) over
   the entire region. We show the location of the mean peak of the isolated 
   component (dashed black line) and the ``mean dip'' (dashed gray line), see 
   text.
   {\bf Middle and right panels:} Normalized example spectra (within a pixel) 
   of the \nhp isolated velocity component extraction procedure 
   (see \S~\ref{sect:iso_extraction}).
   We show G353 \nhp spectra with solid black lines and the extracted
   isolated component, along with emission free channels, with dashed
   red lines.
   {\bf Panel d):} Expected \nhp emission for an excitation temperature of 
   15\,K, an opacity of 1, velocity centroid of $-17\,$\kms, and a line 
   width of 0.2\,\kms. We see the seven hyperfine components characteristic of
   this tracer, where the most blueshifted corresponds to the isolated
   component. To derive this emission we use ``n2hp\_vtau'' model from 
   \texttt{PySpecKit}.   
   In panels b), c), e), and f) we indicate V$_{mean \ dip}$ with a 
   dashed gray line. We present data with SNR~$<$~5 in panel b), 
   where we make a rough extraction based on the V$_{mean \ dip}$.
   We show data with SNR~$\geq$~5 in panels c), e), and f), presenting clear
   single, double, and triple \nhp isolated velocity components respectively.
   In these examples we represent the selected velocity guess that
   separates the isolated component emission from the main line emission
   with dashed blue lines.
   The offset positions ($\Delta l$, $\Delta b$) of the spectra in 
   panels b), c), e), and f) are \text{(0.68\,pc, 0.27\,pc)}, 
   \text{(0.12\,pc, -0.28\,pc)}, \text{(-0.08\,pc, 0.47\,pc)}, 
   \text{(0.16\,pc, -0.07\,pc)} respectively. 
   These offsets are estimated relative to the center of the 
   region (See \S~\ref{sect:intro}).   
   }
   \label{fig:iso_extraction}
\end{figure*}

\section{\nhp isolated component extraction}
\label{sect:iso_extraction}
The \nhp (1$-$0) transition is characterized by its hyperfine 
emission composed by seven components \citep[][see their Fig.~1]{caselli95}.
We present an ideal example of \nhp emission in Fig.~\ref{fig:iso_extraction},
panel d.
In this work we refer to the triplet of hyperfine components
that present the highest intensities as the main \nhp components, 
located at the center of the line emission at $\nu_{rest}~=~93.173806$~GHz. 
We refer to the most blueshifted hyperfine component as the isolated component,
at $\nu_{rest}~=~93.17631$~GHz, shifted by $\sim-8$\,\kms relative to the main 
\nhp component \citep[see Table 1 from ][]{cazzoli85}.
We developed an algorithm to extract only the isolated hyperfine 
component from every pixel in the feathered datacube.
This is in order to reduce the complexity of our data, given that it may
contain multiple velocity components in addition to the hyperfine line
emission.
Considering that the \nhp emission moves in velocity across the protocluster,
our approach is to find the velocity where the emission of the isolated 
component ends and remove the rest of the line emission. We also preserve the 
emission-free channels, at low \text{($-43$\kms~to~$-31.5$\kms)} and high 
\text{($0.7$\kms~to~$6.7$\kms)} velocities, to improve future RMS estimations 
if needed.
Note that in the procedures described below, we use 
\texttt{find$\_$peaks}\footnote{\href{https://docs.scipy.org/doc/scipy/reference/generated/scipy.signal.find_peaks.html}{\url{https://docs.scipy.org/doc/scipy/reference/generated/scipy.signal.find_peaks.html}}}
to detect peaks and valleys in the different spectra.

Our extraction approach is separated into two procedures, 
for low and for high signal-to-noise (SNR) data 
(see Fig.~\ref{fig:snr}, and text below).
In order to determine which data have low or high SNR,
we obtain the mean spectrum over all the spatial pixels of the cube, 
which serves as a guide to determine the velocity at the ``mean dip'' 
(V$_{mean \ dip}$~=~-22\,\kms, dashed gray line in 
Fig.~\ref{fig:iso_extraction} panel ``a'').
This velocity represents the mean location of the intensity valley between
the isolated and the main components of the \nhp emission.
We define $\Delta$V$_{mean}$~=~3.2\,\kms as the difference between
V$_{mean \ dip}$ and the velocity at the peak of the
mean isolated component V$_{mean \ peak}$ (dashed black line in
Fig.~\ref{fig:iso_extraction} panel ``a''), used in our velocity guesses for
the high SNR extraction procedure (see below).

\begin{figure}
\centering
\includegraphics[width=\columnwidth]{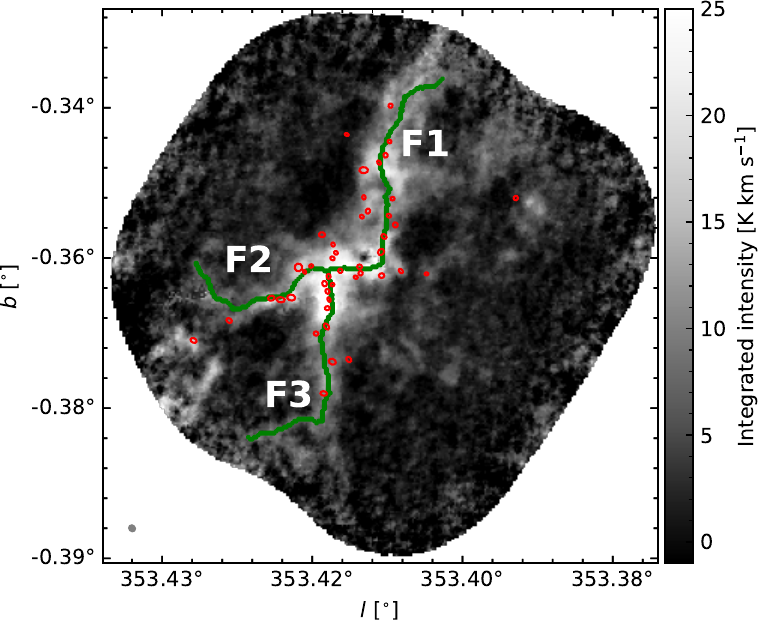}
\caption{Moment 0 map of the extracted \nhp isolated component emission.
We use the \texttt{FilFinder} Python package to identify the main filamentary
structure present in G353 (see Appendix~\ref{app:filfinder}). 
We identify three filaments (F1, F2, and F3; 
green lines) converging towards the central hub. The location of most 
of the 1.3~mm cores (red ellipses), projected in the POS, lie on top 
of the spine of these filaments, specially in the hub.
}
\label{fig:m0_filaments}
\end{figure}

To create a SNR map of the isolated component, we first measure the RMS noise
in emission-free channels \text{($-43$\,\kms~to~$-31.5$\,\kms)}, 
and the peak intensity in the channels range where the mean isolated 
component is located \text{($-~43$\,\kms~to~V$_{mean \ dip}$)}. 
This approach allows us to exclude the emission of the main line components.
We encountered spurious emission at the edges of the SNR map.
We adopt the procedure from \citet[][]{towner24} by using the image
processing techniques implemented by
\texttt{binary$\_$erosion}\footnote{\href{https://docs.scipy.org/doc/scipy/reference/generated/scipy.ndimage.binary_erosion.html}{\url{https://docs.scipy.org/doc/scipy/reference/generated/scipy.ndimage.binary_erosion.html}}} (1 iteration) and
\texttt{binary$\_$propagation}\footnote{\href{https://docs.scipy.org/doc/scipy/reference/generated/scipy.ndimage.binary_propagation.html}{\url{https://docs.scipy.org/doc/scipy/reference/generated/scipy.ndimage.binary_propagation.html}}} to
clean the data for further analysis. \texttt{binary$\_$erosion} allows us to 
remove the spurious emission in the outskirt of the map, although this 
approach also removes high SNR edges of our protocluster. Then, we use \texttt{binary$\_$propagation} on the cleaned SNR map, using the original SNR map mask, to restore only the protocluster edges. 
To test our cleaning approach we compute the total integrated intensity 
using the Python package 
\texttt{SpectralCube}\footnote{\href{https://github.com/radio-astro-tools/spectral-cube}{\url{https://github.com/radio-astro-tools/spectral-cube}}} 
in the range of \text{$-$31.5\,\kms~to~V$_{mean \ dip}$} using the 
original and cleaned SNR mask. We estimate that the removed spurious 
emission accounts for $\sim$2\% of the total integrated intensity for 
data with SNR$>5$.

\begin{figure*}
\centering
\includegraphics[width=\textwidth]{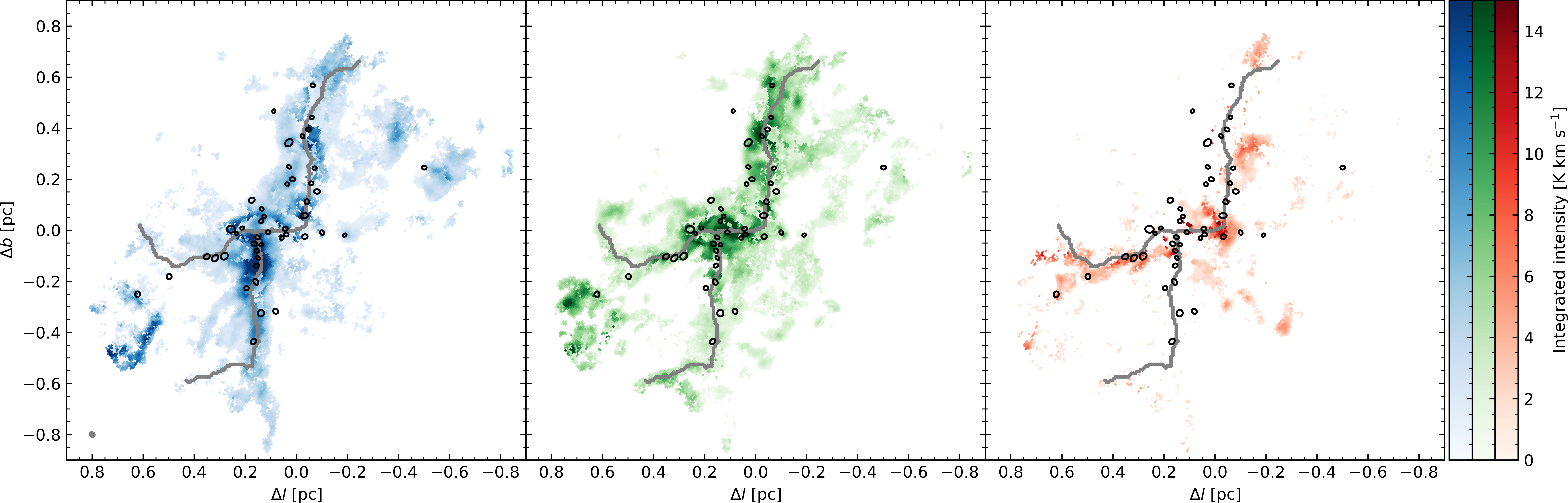}
\caption{
Spatial distribution of the modeled Gaussian velocity components 
that describe the \nhp isolated component emission. We indicate
the main filament structure with green lines (see Fig.~\ref{fig:m0_filaments}).
In blue, green, and red we indicate the first, second, and third velocity
components respectively. We indicate the beam size of these data
with a gray ellipse at the bottom left corner.
The emission of the first and second components is
more extended and intense than the third, most red-shifted velocity
component. The 1.3~mm cores match regions with high integrated intensity,
mostly traced by the first and second velocity components.}
\label{fig:m0_components}
\end{figure*}

In Fig.~\ref{fig:snr} we show the \nhp isolated component SNR map, where
at SNR values $\geq$~5 we capture the cloud emission while excluding 
noise (white contour).
We set our isolated component SNR threshold to 5, in order to use
one of the two extraction procedures (see below). 
In this section we refer to high (low) SNR spectrum if 
its isolated component \text{$\rm{SNR}~\geq~5$ ($<~5$)}.
For low SNR spectra, we extract all the channels in the velocity
range from $-~43$\kms up until V$_{mean \ dip}$ 
(panel ``b'' in Fig.~\ref{fig:iso_extraction}). We take this approach 
given that for low SNR data we can not identify peaks in the \nhp emission 
in a reliable manner.
For high SNR spectra the extraction procedure consists of creating different
velocity guesses that represent the location of the intensity valley,
similar to the definition of V$_{mean \ dip}$ (Fig.~\ref{fig:iso_extraction}).
Then, we select a velocity guess based on its associated weight 
(see description below). This approach is described in detail here:
\begin{itemize}
\item First, we implement a rolling average along each spectrum.
This is in order to smooth over intensity bumps that might result in 
false positives for the detection of peaks and valleys.
For this procedure, we average considering two channels 
before and after each velocity.
\item After smoothing, for each spectrum we identify the 
isolated component peak using \texttt{find\_peaks}. 
We call the velocity associated to this peak V$_{isolated \ component}$.
In the case of multiple velocity components it represents the most 
blueshifted one. We find the intensity valley between the isolated component
and the \nhp main line emission by inverting the spectrum and finding 
the first peak which is the inverted intensity valley. We define the
associated velocity to this intensity valley as V$_{first \ minima}$.
\item We create three velocity guesses based on the properties of 
each spectrum in our cube (see points below). These are the
1$^{st}$ guess: V$_{isolated \ component} + \Delta $V$_{mean}$.
2$^{nd}$ guess: V$_{first \ minima}$,
In the case of multiple isolated components this guess 
might incorrectly capture the intensity valley after the first 
isolated component. In that case, the other guesses are needed for a 
reliable isolated component extraction.
3$^{rd}$ guess: V$_{mean \ dip} + \Delta $V$_{mean}$ 
to provide a velocity cut further away from the V$_{mean \ dip}$.
This guess is mainly useful in the case where multiple 
isolated components cover a velocity range larger than the one 
probed by the other two guesses.
\item From each velocity guess we estimate two parameters to later 
decide which one to use. One is the absolute value of its associated 
intensity ``$I_i$'' (i.e. intensity at the guess velocity), and the other
is distance in velocity ``$dV_i$'' to the mean dip.
The $i$ subscript represents the guess associated to these parameters.
We save the parameters of each guess in the lists ``$I$'' and ``$dV$''.
\item We normalize these lists by their minimum value ensuring that the 
guess with the smallest ``$I_i$'' and ``$dV_i$'' will have a weight ($w$) 
of 1, defined in Eq~\ref{eq:weight}. We do not encounter divergences 
in this normalization given these parameters are not exactly zero. 
\begin{eqnarray}
w = (I_{norm}\times0.2+dV_{norm}\times0.8)^{-1},
\label{eq:weight}
\end{eqnarray}
where the ``norm'' subscript indicates that the parameter list is 
normalized by dividing it by its minimum value.
\item By visual inspection we consider that we obtain good 
extraction results when the weight is mostly dependent on $dV$ and 
in a minor part on $I$. This is reflected by the 0.2 and 0.8 factors 
multiplying $I_{norm}$ and $dV_{norm}$ respectively, in the definition 
of ``$w$'' in Eq~\ref{eq:weight}.
\item We choose the guess with the weight closest to unity.
\item Similarly as for SNR~$<$~5, we extract the spectra from $-~43$\kms up
until the velocity of the chosen guess, and preserving the emission-free 
channels from 0.7\kms to 6.7\kms.
\end{itemize}

Various examples of \nhp spectra and isolated hyperfine component 
extraction are shown in Fig.~\ref{fig:iso_extraction}, where we
can see spectra containing one (panel ``c''), two (panel ``e''), and
three (panel ``f'') velocity components, all well extracted by our 
procedure. In Stutz et al. (in prep) this approach is generalized
to all ALMA-IMF regions for \nhp, providing reliable results. 

\subsection{Filamentary identification}
\label{sect:fil_ident}

We use the \texttt{FilFinder} Python package \citep[][]{filfinder} 
in order to detect the most prominent filaments in this region 
(green lines in Fig.~\ref{fig:m0_filaments}). The procedure, including 
the parameters we used for the filamentary identification, 
is presented in Appendix~\ref{app:filfinder}.
In Fig.~\ref{fig:m0_filaments} we indicate the detected filaments
with green lines, on top of the moment zero map of the multiple \nhp 
isolated components. We see that G353 is a hub-filament system 
(HFS), composed by three main filaments converging at its center.
The HFSs are a characteristic feature of early stages of star formation,
where gas flows through the filaments towards the central hub triggering
star formation
\citep[][]{myers09,galvan10,busquet13,galvan13,peretto14,kumar22,atomsxi}.
We see that in the plane of the sky (POS) most of the 1.3~mm 
cores (red ellipses) are located on top of the filaments.
This spatial agreement between filaments and protostelar cores 
is consistent with filamentary fragmentation 
\citep[][]{andre10,busquet13,stutz15,kuznetsova15,kuznetsova18}.

\section{\nhp isolated component velocity decomposition}
\label{sect:line_modeling}

In Fig.~\ref{fig:iso_extraction} we see that clear multiple isolated 
velocity components are present in our dataset. 
To characterize the complex dense-gas kinematics traced by \nhp we follow the 
method in \citet[][]{alvarez21}, and we use the spectroscopic 
toolkit \texttt{PySpecKit} \citep{ginsburg2011,ginsburg22b} to model and 
decompose the isolated component emission.
\texttt{PySpecKit} adjusts a fixed number of components set by the user,
based on visual inspection of the data we impose three velocity components 
to every spectrum and then remove false positives (see below).
Given the kinematic complexity of the data and cursory inspection of the 
spectra, a simpler analysis with only two components contradicts the data.  
In essence, three components is the simplest possible choice, given the data.  
While this might fail for a small number of spectra that could require 
$\geq4$ velocity components,  the residuals indicate that this could occur 
in a severe minority of cases, and hence more components is not warranted 
given the SNR and resolution of this particular data set. 
To improve the convergence of \texttt{PySpecKit}, we create a set of 
ranges for the
parameters that define each of the three Gaussian velocity components,
namely the peak intensity, central velocity, and velocity dispersion.
After testing different parameter ranges, we set the intensity
range between 1.76\,K (4 times the mean RMS) and 30\,K, the velocity
centroid from $-$30\,\kms to $-$20\,\kms, and the velocity dispersion from 
0.22\,\kms to 1\,\kms.

From the results using the ranges defined above, we notice that some modeled
components do not fit any emission. These fits are the result of imposing
to the fitter a fixed number of components, given these spectra can be better
represented by one or two velocity components.
In these fits there is no uncertainty estimation for both the peak 
intensity and velocity dispersion. Based on these two criteria we remove 
those velocity fits from the modeled cube. 
With this cleaning approach we are left with spectra characterized by 
one ($\sim34\%)$, two ($\sim53\%)$, and three ($\sim13\%)$ Gaussian velocity 
components. We present the Gaussian fits of the high SNR spectra 
from Fig.~\ref{fig:iso_extraction} in Appendix~\ref{app:gaussian_fitting}.

\begin{figure}
\centering
\includegraphics[width=\columnwidth]{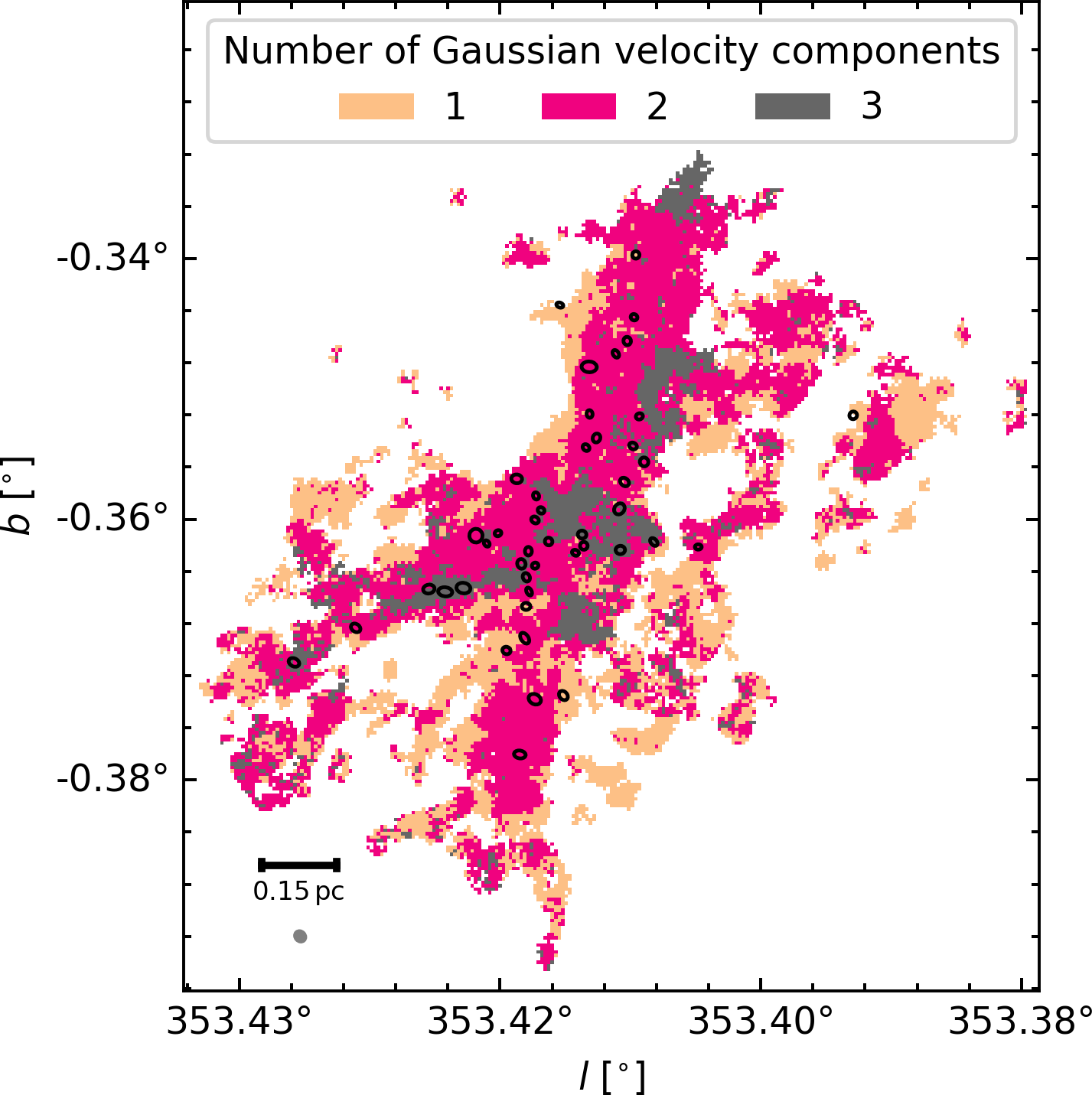}
\caption{Spatial distribution of the \nhp spectra with up to three
   Gaussian velocity components. The 1.3~mm cores and beam size are
   the same as in Fig.~\ref{fig:snr}. Most of the 1.3~mm cores 
   are located in regions with spectra presenting two to three 
   velocity components (see \S~\ref{sect:line_modeling}).
   }
\label{fig:n_comps}
\end{figure}

In Fig.~\ref{fig:m0_components} we show the spatial distribution of the 
multiple Gaussian velocity components. In gray we indicate the main
filamentary structure in the region (see \S~\ref{sect:fil_ident}). 
The first and second velocity components, in blue and green respectively, 
present most of the high intensity emission and they also spatially dominate 
over the third, most red-shifted component. Both the first and second 
components trace mostly the filaments F1 and F3 from 
Fig.~\ref{fig:m0_filaments}, where most of the 1.3~mm 
cores are located. The position of these cores coincide with 
high integrated intensity regions in these isolated velocity components.
The most redshifted component is compact and less intense compared to
the first and second velocity components. This velocity distribution
is located mostly along the filament F2 and the central hub 
(see Fig.~\ref{fig:m0_filaments}). 
In Fig.~\ref{fig:n_comps} we present the number of Gaussian velocity
components for each spectrum, where we highlight that:
\begin{itemize}
\item Most of the \nhp data presents emission characterized by two
velocity components.
\item Most of the spectra described by three velocity
components are located in the innermost parts of the region.
\item Most of the cores (black ellipses; Louvet et al. submitted) are 
located in regions with spectra presenting two to three Gaussian 
velocity components, indicating kinematic complexity even at $\sim$4~kau
(\nhp spatial resolution).
\item Single velocity component spectra are located preferentially in the 
outskirt of the protocluster.
\end{itemize}

\begin{figure}
\centering
\includegraphics[width=\columnwidth]{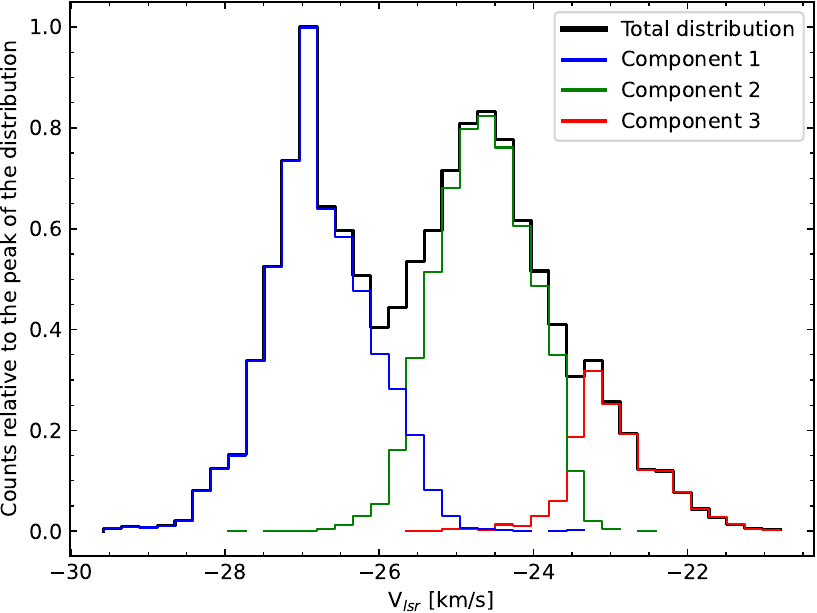}
   \caption{Normalized velocity centroid distributions of each \nhp
   Gaussian velocity component.
   The velocities at the peak of the distributions are $-~26.9$\kms,
   $-~24.7$\kms, and $-~23.3$\kms for component 1 (blue), 2 (green), and 3
   (red) respectively (see \S~\ref{sect:line_modeling}).
   }
   \label{fig:hist_vel_comps}
\end{figure}

In Fig.~\ref{fig:hist_vel_comps} we show the histogram of the fitted velocity
centroid of each Gaussian velocity component. The peaks of these 
distributions are located at $-$27, $-$24.7, and $-$23.3\,\kms respectively, 
well-separated in velocity. From hereafter we refer to these distribution
as blue, green, and red respectively.
Most of the velocity components appear to be associated with the blue
and green distributions.
For consistency with the different tracers used in further analysis,
we shift the isolated component velocities by +8\,\kms, to the reference 
frame of the main line components of \nhp \citep[][]{cazzoli85}.

\subsection{DCN \& \nhp derived core velocities}
\label{sect:nhp_vels}

In this section our goal is to increase the sample of 
core velocities from the already published DCN catalog, aiming to
explore all the potential in these types of dataset.
Given the relatively high $n_{crit}$ of DCN (3$-$2) ($\sim~10^7\,$cm$^{-3}$)  
compared to \nhp (1$-$0) ($2\times10^5~\rm{cm}^{-3}$), DCN (3$-$2) 
is known to coincide well with continuum peaks associated to cores 
\citep[][]{liu2015,cunningham2016,minh2018}, while \nhp is characterized
by tracing the dense gas at the innermost parts of star 
forming regions \citep[][]{fernandez14,hacar18,gonzalez19}. 

In \citet[][]{cunningham2023} they use ALMA-IMF 12~m observations of 
DCN (3$-$2) to study cores kinematics. They apply line emission fits 
for the DCN spectra inside the 1.3~mm cores from Louvet et al. (submitted). 
For this procedure they determine core velocities in all ALMA-IMF targets. 
They classify as DCN single and complex
core velocities, spectra that can be fitted with one or multiple Gaussian
velocity components respectively.
Due to a global conservative SNR threshold the DCN fitting process missed 
the velocity estimation of some cores. For G353 only 15 out of the 45 
cores present DCN velocity fits.

We use the ALMA-IMF DCN 12~m data from \citet[][]{cunningham2023}, which
presents a velocity resolution of $\sim\,0.34$\,\kms.
For each DCN core velocity described by a single component 
\citep[][]{cunningham2023} we compare the emission of the DCN and modeled \nhp 
isolated spectra. 
We find an average velocity offset between the DCN peak and 
the closest \nhp isolated component peak of $\sim\,0.65$\,\kms, less
than two DCN channel widths. We use this approach for the remaining
30 cores, in order to determine their DCN velocities.

Here we estimate the RMS of the DCN data in emission-free channels in the 
range of $-$42\,\kms to $-$25\,\kms and we obtain the SNR map by dividing 
the peak intensity by the RMS. For the procedure below we only use DCN 
spectra with SNR\,$>$\,3. 
Next, we extract the average DCN and modeled \nhp isolated component 
spectrum of these 30 cores. We identify the \nhp isolated velocity 
component closest to the DCN peak within three DCN channel widths.
We find that 11 out of these 30 cores present DCN with SNR\,$>\,3$ close 
to one \nhp velocity component. Here, we define the velocity of these
cores as the velocity where the DCN emission peaks. On average, 
these cores have a velocity offset between these two tracers less than 
0.8\,\kms ($<\,2.5$ DCN channels), similar to the results obtained for 
the 15 cores with DCN velocities from \citet[][]{cunningham2023}, and 
they present an average velocity offset of 0.38\,\kms.  Throughout this 
paper we refer to these cores as ``DCN \& \nhp cores'' given they are derived 
from the comparison of these two tracers.
In Appendix~\ref{app:nhp_core_vels} we present two examples of 
the comparison of the DCN and \nhp spectra in cores where we see clear 
agreement between these tracers (Fig.~\ref{fig:n2hp_cores_vel}).
In Table~\ref{table:core_vels} we show these obtained core velocities from
our comparison between DCN \& \nhp spectra, complementing the DCN 
catalogue from \citet[][]{cunningham2023}.

\begin{figure}
\centering
\includegraphics[width=\columnwidth]{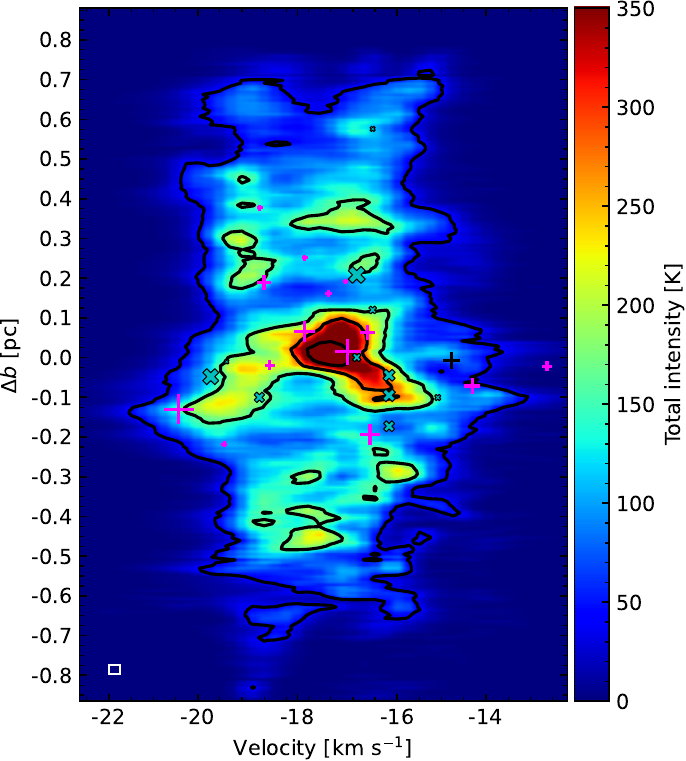}
\caption{``Traditional'' PV diagram of the \nhp modeled isolated components,
created by collapsing the $l$ coordinate. $\Delta b$ indicates the distance
along $b$ in pc, relative to the center of G353, assuming a distance of 
2~kpc \citep[][]{motte2022}. The colormap indicates the total intensity 
along $l$. 
With fuchsia and black crosses we show the 1.3~mm cores with
single and complex DCN velocities detections, respectively 
\citep[][]{cunningham2023}.
With dark cyan ``$\times$'' markers we show the 1.3~mm cores with velocities
derived from DCN and \nhp data (see \S~\ref{sect:nhp_vels}).
The size of the markers indicate relative mass (Louvet et al. submitted).
Black contours indicate total intensities at 40, 160, 280, and 400~K.
We see a large scale velocity spread ($\Delta$V~$\sim$~8\,\kms) around
$\Delta b\sim -0.3~\pc~-~0.13~\pc$ (see also \S~\ref{sect:pv_diagrams}).
We show the major axis of the beam and the channel width with a white 
rectangle at the bottom left corner.
}
\label{fig:traditional_pv}
\end{figure}

\begin{figure*}
\centering
\includegraphics[width=\textwidth]{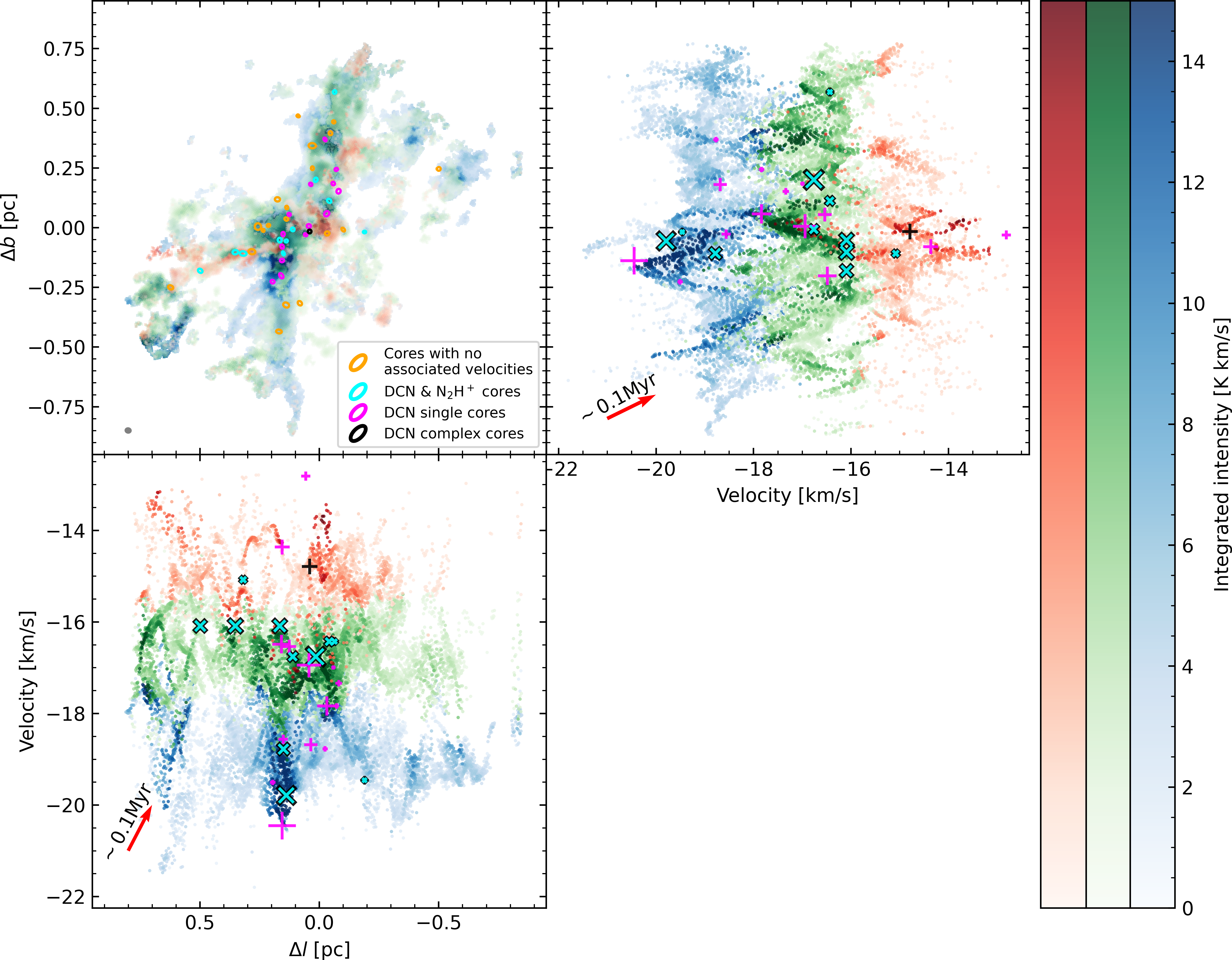}
   \caption{
   {\bf Top Left:} Spatial distribution of the fitted \nhp Gaussian
   isolated velocity components
   (blue, green, and red, see \S~\ref{sect:line_modeling}).
   Ellipses indicate the location of the 1.3~mm continuum cores
   (Louvet et al. submitted). 
   Orange indicates cores with no DCN detections.
   Fuchsia and black represent cores with single and complex DCN velocities
   \citep[][]{cunningham2023}. DCN \& \nhp cores are indicated with cyan. 
   We show the beam size with a gray ellipse in the bottom left corner.
   {\bf Top right and bottom left:} Intensity-weighted position-velocity
   diagrams along the $b$ and $l$ coordinates respectively.
   For the 1.3~mm core velocities we use the same colors
   and markers convention from Fig.~\ref{fig:traditional_pv}.
   For reference we indicate with a red arrow, in both the top right 
   and bottom left panels, a velocity gradient of 10\,\kmspc 
   corresponding to a timescale of $\sim0.1$~Myr.
   We see multiple V-shapes near the location of cores across all
   velocities in the PV diagrams, more prominently in the top right panel
   (see Fig.~\ref{fig:vshapes_app_d}).
   The most prominent V-shape is located in the blue component, at
   $(\rm{V}, \Delta b)~\sim~(-$20.5\kms, $-0.14$\,pc)
   (see \S~\ref{sect:pv_diagrams}).
   We provide an interactive 3D PPV diagram at:
   \href{https://www.rodrigoalvarez.space/research/figures}{\nolinkurl{rodrigoalvarez.space/research/figures}}.
   }
\label{fig:panel_pv}
\end{figure*}

\section{Analysis of position-velocity diagrams}
\label{sect:analysis}

\subsection{Traditional PV diagram}
\label{sect:pv_trad}

We start by analyzing the ``traditional'' PV
diagram shown in Fig.~\ref{fig:traditional_pv}. 
We create this diagram by taking the total intensity along 
the Galactic longitude, where $\Delta b$ indicates the distance 
in parsec relative to the center of G353, assuming a distance to the
protocluster of of 2~kpc \citep[][]{motte2022}.
We see general agreement between the DCN core velocities and the \nhp
velocity distribution. This suggests that most of the cores are still
kinematically coupled to the dense gas in which they formed.
As presented in \S~\ref{sect:nhp_vels}, the DCN and \nhp velocities
match within 0.8\,\kms ($<\,2.5$ DCN channels).

\begin{figure}
\setlength{\tabcolsep}{1pt}
\begin{tabular}{c}
\begin{overpic}[width = \columnwidth]{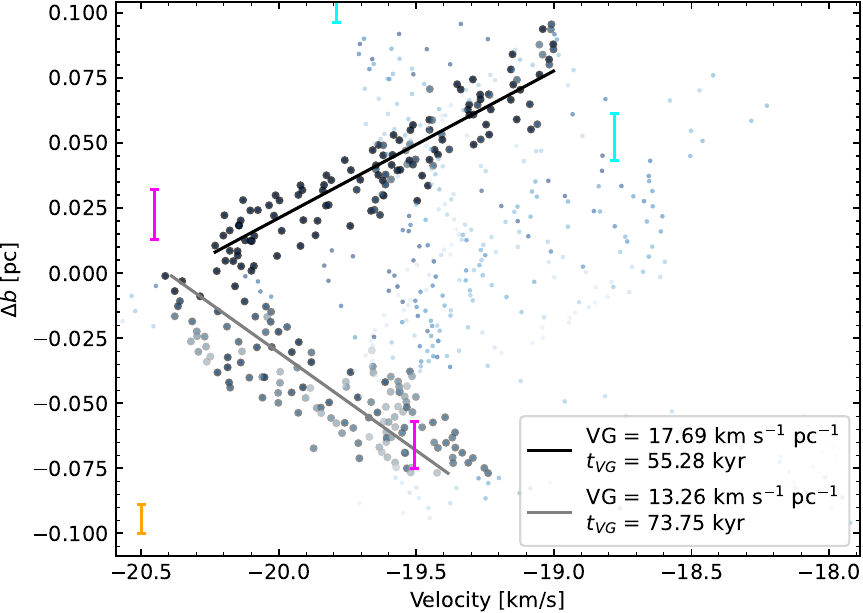}
\put(15,66){\large C} \end{overpic} \\
\end{tabular}
\caption{Zoomed-in version of the top right panel of Fig.~\ref{fig:panel_pv},
centered at the prominent blue V-shape (``C'') located at
$(\Delta b, \rm{V})~=~(-0.14$\,pc, -20.5\kms).
We indicate the major axis of the cores with vertical lines in 
fuchsia (DCN single) and cyan (\nhp), similarly we represent
the major axis of the beam with an orange vertical line.
We apply linear fits to the upper and lower distributions, 
represented by darker points. These points are selected based on an integrated
intensity threshold (see \S~\ref{sect:vel_grad}).
We weight each point by their integrated intensity and derive
velocity gradients (VGs) from the slope of these linear fits.
The range of the obtained VGs is $\sim~13-18$\kmspc.
We define the timescales associated to the VG as $t_{VG}~=~\rm{VG}^{-1}$,
being in the range of $\sim50-70$\,kyr. We show eight more well 
characterized V-shapes in Appendix~\ref{app:v-shapes}.
}
\label{fig:main_gradient_blue}
\end{figure}
Regarding the dense gas velocity distribution, in 
Fig.~\ref{fig:traditional_pv} we see a velocity spread of $\sim~8~$\kms
in the sub-region between \text{$\Delta b\sim$~$-$0.3~pc~to~0.1~pc}.
Most of the intensity on this diagram is located at the upper part of
this sub-region, at $\Delta b~\pm~0.1~\rm{pc}$. This spread is also present in
the PV diagram along $\Delta b$ and $\Delta l$ shown in the top right
and bottom left panel of Fig.~\ref{fig:panel_pv}. We explore the possible 
origin of this structure in \S~\ref{sect:collapse}.

\subsection{Intensity-weighted PV diagrams}
\label{sect:pv_diagrams}
In the top left panel of Fig.~\ref{fig:panel_pv} we show the spatial
distribution of the fitted Gaussian velocity components
(see \S~\ref{sect:line_modeling}).
The blue, green, and red color maps indicate the integrated intensity
of the first, second, and third velocity components of the \nhp spectra
respectively. Note that the spatial overlap between any of these components 
is presented in Fig.~\ref{fig:n_comps}.

\begin{figure*}[!h]
\centering
\includegraphics[width=\textwidth]{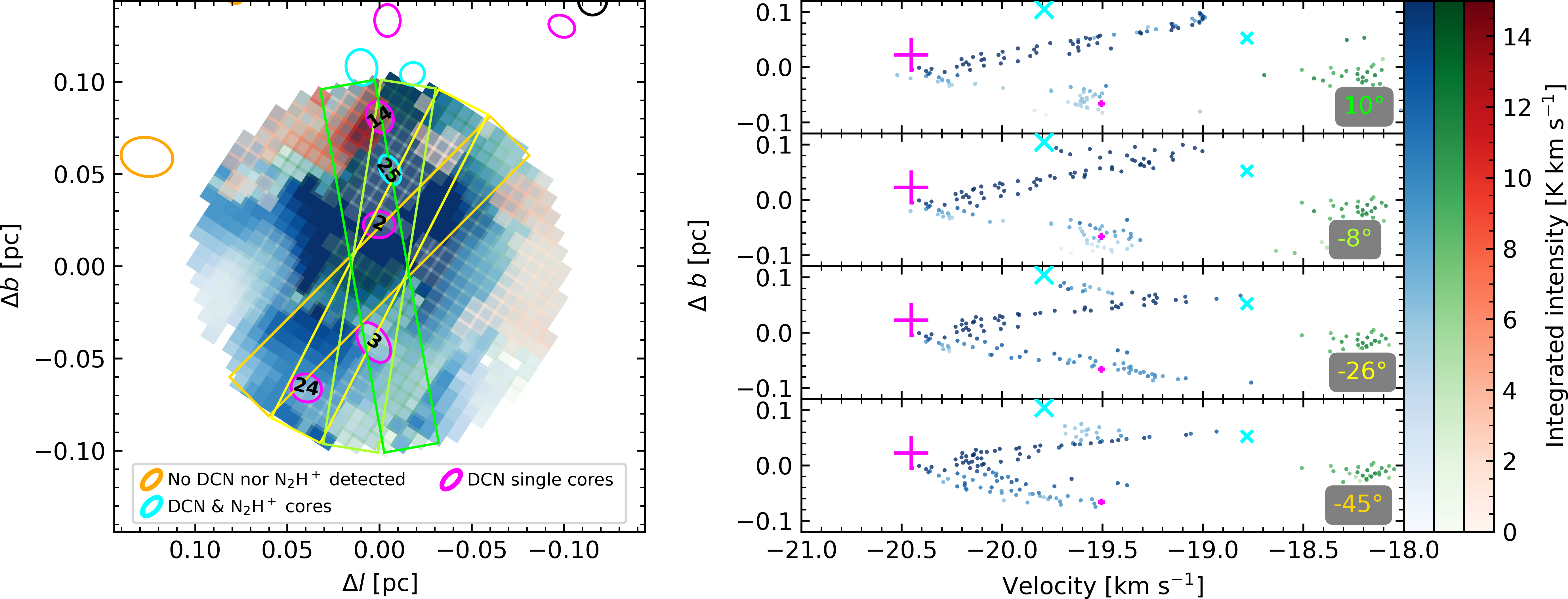}
   \caption{{\bf Left panel:} Integrated intensity map of the modeled
   \nhp isolated component, inside a region centered at the main blue V-shape
   with a radius of 0.1\,pc. With boxes in shades of green and
   yellow, we indicate the different paths taken to create the PV diagrams
   shown on the right panel.
   The area covered by these four PVs matches the extent of this V-shape 
   (see Fig.~\ref{fig:multi_tracer}).
   The 1.3~mm cores are indicated using the same convention from 
   Fig.~\ref{fig:panel_pv}. For the cores within the radius of 0.1\,pc
   we indicate their IDs in black.
   {\bf Right panel:} PV diagrams corresponding to the different paths
   presented in the left panel. At the bottom right corner
   of each sub panel, with colored values, we indicate the angle
   (counter-clockwise) of each PV path.
   The PV paths match the V-shape coverage
   (right panel of Fig.~\ref{fig:multi_tracer}). We see the overall
   structure of this V-shape persists at different angles, indicating that
   these structures are not a result of projection in the POS.
   }
   \label{fig:pv_angles}
\end{figure*}
As seen in Fig.~\ref{fig:traditional_pv}, the traditional PV diagram 
provides information on the dynamics on the large, protocluster-scale,
environment. Meanwhile, the intensity-weighted position-velocity diagram
(Fig.~\ref{fig:panel_pv}), where the color of each point indicates its
integrated intensity, highlights the small core-scale kinematics.
Similarly as in \citet[][]{gonzalez19} and \citet[][]{alvarez21}, 
from the isolated component line decomposition 
(\S~\ref{sect:line_modeling}), we derive the integrated
intensity and velocity centroid for each Gaussian velocity component.
Using these parameters we create intensity-weighted PV
diagrams along the $b$ and $l$ coordinates.
We present these \nhp PV diagrams in the bottom left and top right
panels of Fig.~\ref{fig:panel_pv}. 
The key features on the position-position (PP) and on 
the top right PV diagram, are:
\begin{itemize}
\item The agreement between the DCN core velocities and the overall \nhp
PV structures suggests that cores are still kinematically coupled
to the dense gas in which they formed.
\item We see at least nine clear and prominent V-shaped velocity
gradients (see Appendix~\ref{app:v-shapes}), across all 
velocity components. The orientation of these
V-shape, pointing to the left/right (top right panel) or up/down (bottom left
panel), follow no clear preference.
\item  In some cases, the vertex of these V-shapes is 
close spatially and in velocity to the location of cores.
\item In the plane of the sky (POS), all three velocity components overlap
in most of the region.
\item This technique recovers the large scale velocity spread present 
in Fig.~\ref{fig:traditional_pv} and highlights small scale structures.
\item The most prominent V-shape is located at 
\text{$(\Delta b, \rm{V})=(-0.14$\,pc, $-$20.5\,\kms)}, between 
two 1.3~mm cores with DCN detections (see \S~\ref{sect:collapse}).
\end{itemize}

For a better visualization of the 3D structure of these V-shapes we
provide an interactive 3D PPV diagram at:
\href[pdfnewwindow=true]{https://www.rodrigoalvarez.space/research/figures}{\nolinkurl{rodrigoalvarez.space/research/figures}}.

\begin{figure*}[]
\centering
\includegraphics[width=\textwidth]{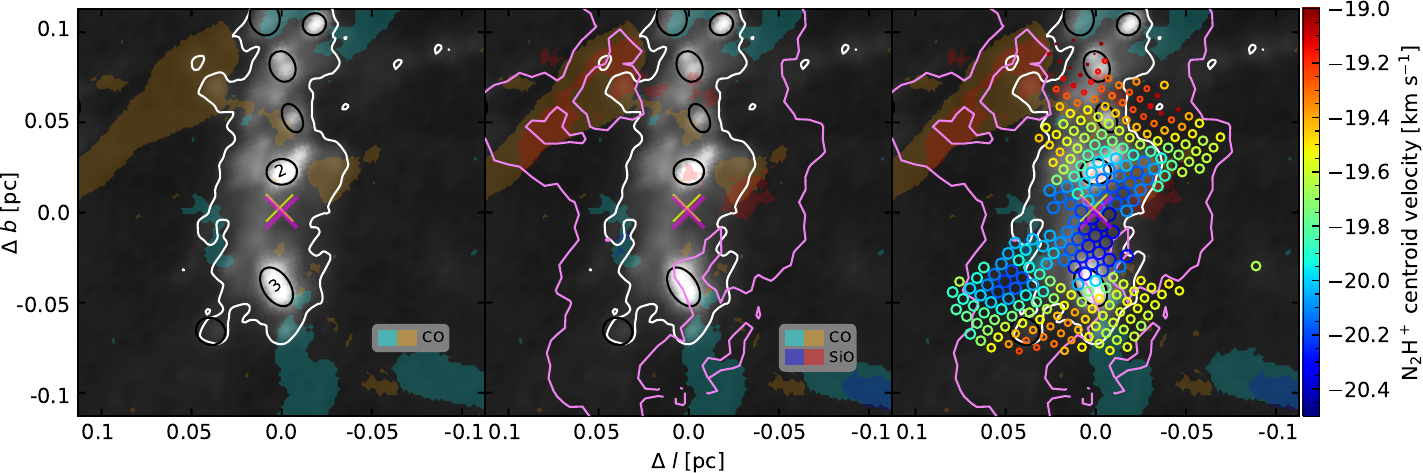}
   \caption{Multi-tracer diagram at the location of the main blue
   V-shape of G353 (\S~\ref{sect:vel_grad}). The different panels show
   the step by step construction of the final plot (right panel).
   {\bf Left panel:} The black to white background shows the 1.3~mm 
   continuum emission from \citet[][]{diaz23}.
   With black ellipses we show the 1.3~mm continuum cores from Louvet
   et al. (submitted). With black text we indicate the IDs of the cores
   closest to the center of the V-shape, marked with a fuchsia ``$\times$''.
   We indicate the barycenter of cores 2 \& 3 (see Fig.~\ref{fig:pv_angles}) 
   with a yellow ``$\times$''.
   Filled contours represent the CO (2-1) emission in the velocity
   ranges of \text{$-50$\,\kms to $~-15$\,\kms} (cyan) and 
   \text{$15$\,\kms to $50$\,\kms} (beige), relative to the 
   V$_{LSR}~=~-17$\kms of G353.
   With a white contour we indicate the 
   \text{$log(N{\rm(H}_2) \ \rm{cm}^{2})~=~23.3$}.
   {\bf Middle panel:} 
   In addition to the left panel, we include the SiO ($5-4$) emission 
   with blue and red contours in the same (negative and positive)
   velocity range as with CO. The pink contour indicates the
   integrated intensity of the most blueshifted modeled \nhp Gaussian 
   velocity component (see Fig.~\ref{fig:panel_pv}; blue distribution), 
   at a value of 7\,K\,\kms.
   {\bf Right panel:} With open circles we show the location of the data
   composing the main \nhp blue V-shape. The colors indicate their
   velocity centroid, and their (increasing) size indicates how 
   close the gas velocities are to the velocity apex of the V-shape.
   This velocity gradient seems to converge to the barycenter of
   cores 2 \& 3, and it is oriented along a filament.
   }
   \label{fig:multi_tracer}
\end{figure*}

\subsection{Velocity gradients}
\label{sect:vel_grad}

In this section we focus on the most prominent blue V-shape 
(Fig.~\ref{fig:panel_pv}, top right panel).
In Fig.~\ref{fig:main_gradient_blue} we show this velocity distribution
in detail. 
In order to characterize the VGs composing this V-shape, 
we apply a linear fit to both the upper and lower VG.
Given the visual linearity of the VGs composing the V-shape, we
apply a linear fit to these distribution in order to characterize them.
For these fits we consider data only above an integrated
intensity threshold of 8~K\,\kms and 3~K\,\kms, for the upper and
lower gradient respectively. We remove data
not related to the velocity gradient, clustered in the ranges of
\text{$(\Delta b,\rm{V})~\sim~(-0.025~-~0.04\,\pc~,~-19.5~-~-18.5\,\rm {km\,s}^{-1})$,}
which lie just outside the filament hosting this V-shape on the POS.
Additionally, we weight each point based on their integrated intensity
to make our fits more robust.
The slopes of the linear fits represent the VGs in \kmspc.
These linear fits follow the VGs distribution and these are somewhat
asymmetric, the upper gradient is slightly shallower than the
bottom gradient. 
Given the unknown inclination angle ($\theta$) of these structures 
relative to the POS, the observed VG is just a fraction of the original VG. 
These are related as VG~=~VG$_{original}\cdot\sin(\theta)$.
These VGs present values between \text{$\sim13~$to$~\sim18$\,\kmspc} 
(see Fig.~\ref{fig:main_gradient_blue}).
Additionally, we estimate the center of this V-shape as the
velocity-weighted mean position of the points composing this structure.
With this approach the position of the points closest to the V-shape apex 
present more weight, obtaining the center of this V-shape at
($l,b$)~=~(353.4135\degree, $-$0.3657\degree).
This position is located between ``core 2'' and ``core 3'', 
both of them having DCN velocity fits \citep[][]{cunningham2023}. These
core present masses of 20.7 and 6.4\,\msun respectively. 
We inspect the core catalog derived from the map at native resolution 
(Louvet et al. submitted) and the location of this V-shape do not coincide 
with any core.

In the left panel of Fig.~\ref{fig:pv_angles} we show the 
integrated intensity of the multiple modeled \nhp isolated components.
With colored boxes we show the areas where we create the different
PV diagrams presented on the right panel. These boxes are centered at the main 
blue V-shape, matching the area of this V-shaped structure 
(see Fig.~\ref{fig:multi_tracer}). We show that the overall 
structure in PV space is conserved at different angles, excluding the 
possibility of this velocity feature being the result of projection effects.

\begin{figure}
\centering
\includegraphics[width=\columnwidth]{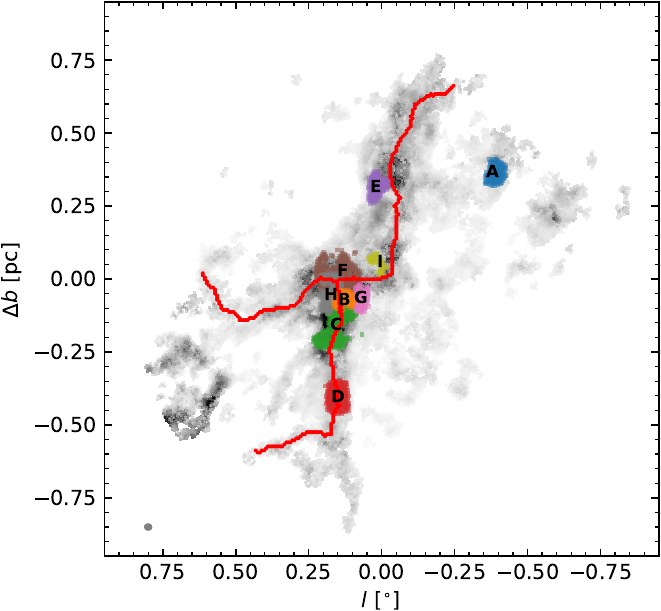}
   \caption{
   V-shapes location on the plane of the sky.
   We indicate the position of each V-shape with colored points and
   their ID with black text. With red lines we show the filamentary 
   structure identified in \S~\ref{sect:fil_ident}.
   In the background we show the integrated intensity map of all 
   three velocity components, similarly as in Fig.~\ref{fig:panel_pv}.
   We see most of of the V-shapes are located at the hub.      
   }
   \label{fig:vshapes_pos}
\end{figure}

\iffalse
For this V-shaped structure, we note that the \nhp velocity range is
\dv~$\sim$~2\,\kms, while the whole \nhp isolated component dataset
present a \dv~$\sim$~8\,\kms (Fig.~\ref{fig:panel_pv}).
\fi
For this V-shape we use its composing VGs to derive timescales as 
$t_{VG}$~=~1/VG, similar to the procedure for a rotating filament 
presented in \citet[][]{alvarez21}, 
and we suggest these can be interpreted as inflow timescales.
The $t_{VG}$ values for this V-shape range between $\sim~50-70$\,kyr. 
These timescales are short compared to the $\sim$~0.21\,Myr free fall time
(t$_{ff}$) of the protocluster \citep[][]{motte2022}, and a few times larger 
than the t$_{ff}$ of nearby cores ($\sim$~20\,kyr, within 0.1\,pc 
of this V-shape). To determine the cores t$_{ff}$, we use the
1.3~mm core masses from Louvet et al. (submitted).

We characterized eight more \nhp V-shaped structures.
In Fig.~\ref{fig:vshapes_pos} we show the position of these V-shapes
in the POS. 
Previous studies have detected velocity gradients along filaments,
towards a converging point \citep[][]{peretto14,pan24,rawat24}.
In the case of G353 we see that instead of detecting a single V-shape
at the intersection of its filaments, the hub appears to be fragmented 
into multiple, small-scale V-shaped VG. Only V-shapes ``A'', ``D'', 
and ``E'' are outside of the hub, with V-shape ``D'' located on top of 
filament ``F3''. This indicates that the V-shaped structures are not 
exclusive to the central parts of HFS, but are also present in comparatively 
isolated regions.
Note that within a $\sim$beam size from 
the apex of V-shape ``B'' (see Fig.~\ref{fig:v-shapes}), the continuum 
core ``7'' ($\sim6\,$\msun) is located.

In \citet[][]{henshaw14}, they propose two scenarios that might produce
these V-shaped velocity gradients (see their Fig.~12). One scenario
suggests that gas in a filament is flowing towards a denser region 
(infall), while the other scenario suggests that a protostellar 
outflow moves the dense gas located in its vicinity.
To analyze the different dynamical processes present in this region, we
use the ALMA-IMF 12~m data of the shock, outflow tracer SiO (5$-$4), from
\citet[][]{cunningham2023}. From those data we create its 
intensity-weighted PV diagram, presented in Fig.~\ref{fig:sio_panel_pv} 
(see Appendix~\ref{app:sio_pv} for more details).
We note that there is almost no SiO emission nor outflow sources at the
location of the \nhp V-shape (see Fig.~\ref{fig:multi_tracer}).
The \dv \ within 10$''$ from this main blue V-shape is $\sim$40\,\kms, 
while for the whole SiO dataset is $\sim$80\kms, showing a clear 
difference in the SiO \dv \ and the core velocities. 
Furthermore, the SiO \dv is $\sim10$ times larger than the one of \nhp.
This difference in traced velocities implies that these two molecules 
trace vastly different physical phenomena. 
We suggest that the small velocity range probed by \nhp indicates that the
velocity gradients can be considered as infall signatures 
(see \S~\ref{sect:collapse}). We discuss the possible morphology of
the filaments hosting V-shapes in \S~\ref{sect:discuss}.

\section{G353 as a collapsing region}
\label{sect:collapse}
We use the SiO ($5-4$) and CO ($2-1$) data from \citet[][]{towner24} and
\citet[][]{cunningham2023} to identify possible outflows near the V-shape 
presented in Fig.~\ref{fig:main_gradient_blue}. For CO we measure the RMS 
in the emission-free velocity range of 145~to~286~\kms. We use only CO
data with SNR$~>~$3 for our analysis. The cleaning of the SiO data is
described in Appendix~\ref{app:sio_pv}.
In Fig.~\ref{fig:multi_tracer}, with a fuchsia ``$\times$'' we
indicate the center of the V-shape from \S~\ref{sect:vel_grad},
located between two 1.3~mm cores 2 \& 3 (black ellipses), which present
DCN velocity detections.
From these diagrams we see there is neither SiO nor CO outflow detection at the
location of the main blue V-shape ($\Delta l$,~$\Delta b$)~=~(0\,pc,~0\,pc).
In the right panel we show the position of the data composing this V-shape, 
where the velocity peaks towards the center of this velocity feature.

We derive the mass-weighted mean position of the two cores (2 \& 3) closest 
to the V-shape to determine their barycenter 
(yellow ``$\times$'' in Fig.~\ref{fig:multi_tracer}). We find that 
the difference between the center of the V-shape and the barycenter of 
these cores is $\sim0.3$\arcsec ($\sim600$au), well below the beam size of 
the \nhp data. A similar offset is also present in the intensity and 
velocity profiles along filaments from ATOMS data 
\citep[][see their Fig.~6]{atomsxi}.
This small spatial offset might suggest that the gas flowing in the V-shape 
is produced by the gravitational pull towards the barycenter of cores 2 \& 3, 
where the \nhp radial velocities peak. This interpretation is similar to
the one proposed in \citet[][]{zhou23} in their kinematic analysis 
of the G333 complex. They describe the V-shaped velocity gradients as 
the result of gas funneling from the molecular cloud to clumps which is then
funneled into cores (see their Fig.~9) consistent with gravitational 
acceleration.

\begin{figure}
\centering
\includegraphics[width=\columnwidth]{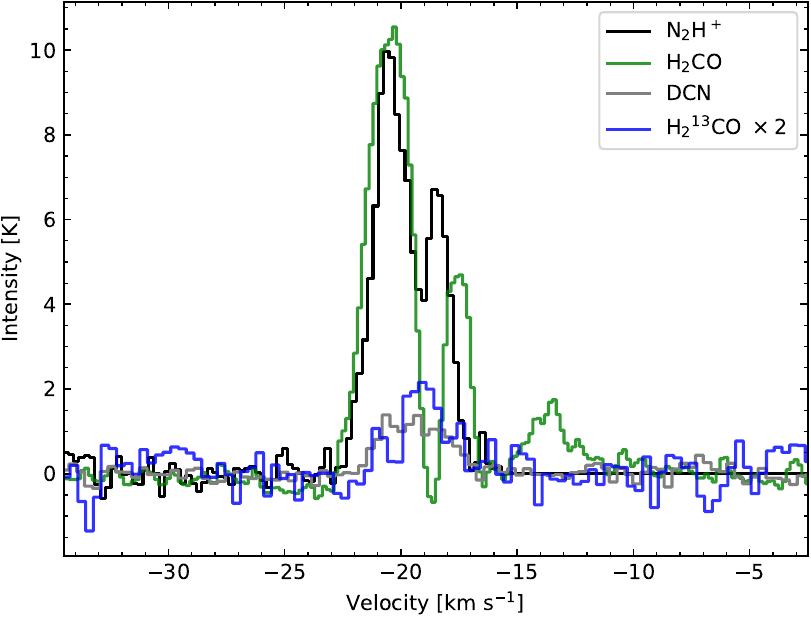}
   \caption{Mean spectra within a 1.14\arcsec \ ($\sim$~0.01~\pc) 
   radius around the location of the main blue V-shape 
   (pink ``$\times$'' in Fig.~\ref{fig:multi_tracer}). 
   Both \nhp and H$_2$CO show blue asymmetry, known to characterize 
   infall motions. The difference in velocity between the two
   \nhp peaks is $\sim\,2.5\,$\kms.
   }
   \label{fig:mean_spectra}
\end{figure}

In Fig.~\ref{fig:mean_spectra} we show the mean spectrum of different
tracers at the central position of the main blue V-shape. These spectra
are measured over a circular region with diameter equal to the 
major axis of the \nhp beam (2.28\,\arcsec; $\sim~0.02~\pc$). 
This circular area results in a coverage of 1.14 times the \nhp beam,
and $\sim$5.6 times the beams of the H$_2$CO, DCN, and H$_2$$^{13}$CO
data. We see \nhp and H$_2$CO present
double component spectrum with asymmetric peaks. Between these peaks
we detect DCN and H$_2$$^{13}$CO emission. 
The asymmetric spectrum present in \nhp and H$_2$CO is consistent
with the ``blue asymmetry'' spectral feature, usually interpreted as 
infall signature, suggesting that this region is under gravitational collapse
\citep[e.g.][]{anglada87,mardones97,lee99,lee2001,rowan2012}. 
Based on the idea that the V-shapes are the result of flowing gas 
along filaments towards denser regions, the blue-asymmetry detected 
at the center of the main blue V-shape suggests that gravitational 
collapse is taking place at the apex of the V-shaped structure.

Regarding large scales, in the traditional PV diagram presented \S~\ref{sect:pv_trad} (see Fig.~\ref{fig:traditional_pv}),
we see a clear velocity spread around $\Delta~b~\sim~-0.1$\,pc, also 
present in the top right panel of Fig.~\ref{fig:panel_pv}. 
Below we compare this velocity spread with the velocity
distribution produced by infall, where the gas velocities increase as 
the distance to the center of infall (``$r$'') decreases:
\begin{eqnarray}
V_{infall} & = & -\sqrt{\frac{2{G\,M}}{r}} 
\label{eq:v_infall}
\end{eqnarray}
%below new model text
For this comparison we model a sphere with a total mass of 150~\msun, 
a radius of 0.5\,pc, and a power law density profile described 
by:
\begin{eqnarray}
\rho(r) & = & \rho_0 \left (\frac{r}{\rm{pc}} \right )^{-\gamma}, \ \
\gamma~=~5.65, \ \ \rho_0~=~6.1\times10^{-5}~\frac{\rm{M}_{\odot}}{\rm{pc}^3},
\label{eq:density_prof}
\end{eqnarray}
we provide the derivation of $\rho(r)$ in Appendix~\ref{app:density_profile}.
$\gamma\,=\,5.65$ was determined by visual inspection by comparing 
the obtained radial velocities of the model (see below), at different
$\gamma$ values, with the overall shape of the PV distribution.

We then estimate the infall velocity of each point, based on the 
cumulative mass distribution (``M'') at any given distance
to the center (Eq.~\ref{eq:v_infall}).
We obtain the radial component of the infalling velocities as:
\begin{eqnarray}
V_r & = & V_{infall}\,\times\,\cos(\arctan(X/Z)),
\label{eq:v_r}
\end{eqnarray}
where $X$ represents the horizontal coordinate in the POS, while
$Z$ represents the (non-observed) depth of the sphere.
The spatial coordinates for this model range from 
$-0.5\,$pc~to~$0.5$\,pc.

\begin{figure}
\centering
\includegraphics[width=\columnwidth]{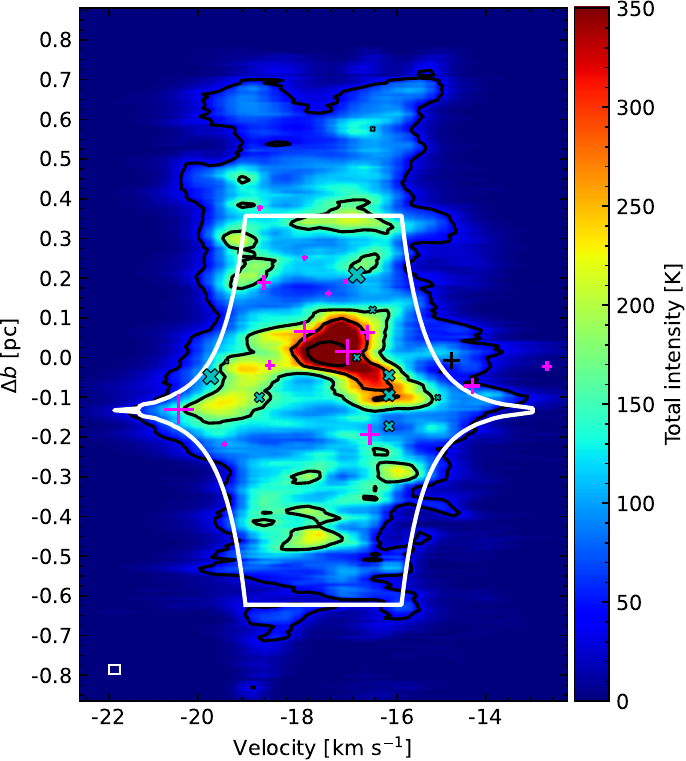}
   \caption{PV coverage of a gravitationally collapsing sphere.
   The white contour represents the coverage of the synthetic radial
   velocities derived from this model (\S~\ref{sect:collapse}).
   The background and cores are the same as in 
   Fig.~\ref{fig:traditional_pv}. For the modeled sphere we set 
   its total mass to 150~\msun, within a radius of 0.5\,pc.
   }
   \label{fig:pv_sphere}
\end{figure}

In Fig.~\ref{fig:pv_sphere}, 
we show the coverage of the PV distribution from our model, described
in Eqs.~\ref{eq:v_infall}\,$-$\,\ref{eq:v_r}, with a solid white line. 
We find good agreement between the 
PV distributions of our approach and the data. 
The PV distribution of our infall model is consistent with previous 
work that provide the expected PV distributions for spherical 
protostellar envelopes under infall \citep[][]{tobin12a}.
At large scales we interpret the agreement between the PV diagrams 
of our model and the data as protocluster scale collapse due to 
gravitational contraction. 
It is worth noting that the inferred mass from our model is 
$\sim$5.5 times lower than the one derived from the $N$(H$_2$) 
map \citep[][]{diaz23}.
We speculate that a model considering complex processes such as  
turbulence, radiative transfer, and magnetic fields might solve
this mass discrepancy while still matching the observed PV distribution.

\section{Mass accretion rates in the V-shaped structure}
\label{sect:mrates}

Based on the idea that the V-shapes are a result of gas flowing 
toward cores, in this section we provide estimates of their 
mass accretion rates (\mrate) for \nhp and H$_2$.

\subsection{H$_2$ mass accretion rate}

To ensure that we estimate the V-shape $M$(H$_2$) on the same area
as in the \nhp hyperfine line fitting, we use the \texttt{CASA} task
\texttt{imregrid} to obtain the continuum-derived $N$(H$_2$) map from
\citet[][]{diaz23} at the resolution of the \nhp data.
We determine that in this V-shape the total
$N$(H$_2)~\sim$~1.17$\,\times\,10^{26}$~cm$^{-2}$ in an area of
0.013\,pc$^2$.
Here, we derive a $M$(H$_2$) map using Eq.~\ref{eq:nh_to_mass}:
\begin{eqnarray}
M\rm{(H}_2) & = & 2 \times  N\rm{(H_2)} \times \rm{area_{pixel}} \times
m_{proton},
\label{eq:nh_to_mass}
\end{eqnarray}
where from this $M$(H$_2$) map we consider only the points that
are part of the V-shape. To determine the mass associated to flowing
motions we subtract the core masses (from Louvet et al. submitted) that
are located inside this V-shape.
Note here that this mass map is an upper limit given that we do not
apply a background correction. We obtain a total of
$M$(H$_2)~\sim$~53~\msun.
Considering $t_{VG \ mean}$ used in Eq.~\ref{eq:mrate_n2hp},
we derive the \mrate(H$_2$) as:
\begin{eqnarray}
\dot{M}_{\mathrm in}({\rm H}_2) \ = \ \frac{M\rm{(H}_2)}{t_{VG \ mean}} \ = \ 8.22\times 10^{-4}~\msun~\rm{yr}^{-1}.
\end{eqnarray}

We use the procedure described in this section to estimate the
\mrate(H$_2$) of other eight V-shapes shown in Fig.~\ref{fig:v-shapes}.
We include these values in Table~\ref{table:v-shapes}.
The average \mrate(H$_2$) of these V-shapes is
3.4\,$\times$\,10$^{-4}$\,M$_{\odot}$ yr$^{-1}$.
Note that V-shapes ``H'', ``C'', ``F'', and  ``B'' present
the largest \mrate(H$_2$) and they are located near or at the
convergence point of the filaments (see Fig.~\ref{fig:vshapes_pos})

For comparison, using the core masses from Louvet et al. (submitted)
we estimate the free-fall time of all 45 1.3~mm cores and their
mass accretion rates. These values present large scattering,
ranging from \text{$(0.07-25)\,\times\,10^{-4}$\,M$_{\odot}$ yr$^{-1}$},
with 28 of these cores presenting
\text{\mrate(H$_2$)\,$<\,10^{-4}$\,M$_{\odot}$ yr$^{-1}$}.
For cores 2 \& 3, the average \mrate(H$_2$) is
\text{15.5$\,\times\,10^{-4}$\,\msun yr$^{-1}$}, about twice the
\mrate(H$_2$) of the main blue V-shape, located between these two cores.

\subsection{\nhp mass accretion rate and relative abundance}

To derive the \mrate associated to the main blue V-shape
(Fig.~\ref{fig:main_gradient_blue}), we need to estimate its \nhp mass. 
For this procedure we use \texttt{PySpecKit} with the \texttt{n2hp\_vtau} 
fitter, to fit the full \nhp line. The fitted parameters (see below) allow
us to derive the $N$(\nhp).
The V-shape structure contains 77, 143, and 25 \nhp
spectra with one, two, and three velocity components respectively.
The bluest velocities in the three velocity component spectra
accounts for $\sim$2\% of the total number of velocities in this V-shape.
For this reason, we model the full \nhp hyperfine line structure 
with one and two velocity components. 
In Table~\ref{table:n2hp_full} we list the parameters and ranges
used for this procedure.

After obtaining the modeled \nhp cube, we remove modeled components 
where $\tau_{\rm{RMS}}/\tau~>~0.3$, where $\tau$ and $\tau_{\rm{RMS}}$ 
represent the estimated opacity and its associated error, respectively. 
This criterion is to ensure that we use reliable fitted parameters
to determine our $N$(\nhp) values. For the fitting of two \nhp components
we only analyze the most blueshifted component. The resulting opacities
follow a log-normal distribution with a peak at $\tau~=~0.13$.

We derive the $N$(\nhp) of the V-shape by using Eq.~\ref{eq:nh_n2hp} \citep[][]{caselli2002b}:
\begin{equation}
N(\mathrm{N}_2\mathrm{H}^+) = \frac{8 \pi^{3/2} \Delta v}{2\sqrt{\ln 2} \lambda^3 A}
\frac{g_l}{g_u} \frac{\tau}{1-\exp{(-h \nu/k \rm{T}_{\mathrm{ex}}})}
\frac{Q_{\mathrm{rot}}}{g_l \exp(-E_l/k\mathrm{T}_{\mathrm{ex}})},
\label{eq:nh_n2hp}
\end{equation}
Where $\tau$, T$_{\rm{ex}}$, and $\Delta v$ are the opacity, excitation
temperature, and velocity dispersion respectively, obtained from the full line 
fitting. The Planck and Boltzmann constants are represented by $h$ and 
$k$ respectively, $\nu$ and $\lambda$ are the frequency and 
wavelength of \nhp,
$A$ is the Einstein coefficient of the \nhp (1$-$0) transition,
$g_l$ and $g_u$ are the statistical weights of the lower and upper 
energy levels,  $E_l$ is the energy of the lower level, 
$Q_{\rm{rot}}$ is the partition function estimated 
using the excitation temperature of the full \nhp fits 
\citep[see Eq. A2 from ][]{caselli2002b}.

From the above procedure, inside the V-shaped structure, we get a total
$N$(\nhp)~=~5.24$\,\times\,$10$^{15}$~cm$^{-2}$ and a total
$M$(\nhp)~=~5.9$\,\times\,$10$^{-8}$~\msun, within its extent of
0.013\,pc$^2$.
We use the average timescale of the VGs from \S~\ref{sect:vel_grad}
(Fig.~\ref{fig:main_gradient_blue}), $t_{VG \ mean}$~=~64.5~kyr,
to determine the \text{\mrate(\nhp)} as:
\begin{eqnarray}
\dot{M}_{\rm in}({\rm N}_2{\rm H}^+) & = &
\frac{M\mathrm{(N}_2\mathrm{H}^+)}{t_{VG \ mean}}
\ \ = \ \ 9.1\,\times\,10^{-13}~\msun\,\rm{yr}^{-1}.
\label{eq:mrate_n2hp}
\end{eqnarray}

Note that the $\dot{M}_{{\rm in}}$(\nhp) estimate (and 
$\dot{M}_{{\rm in}}$(H$_2$) below) should be multiplied by 
$\sin(\theta)$, in order to account for the unknown inclination 
angle ($\theta$) of the protocluster/filaments relative to the POS.

For the main blue V-shape (Fig.~\ref{app:v-shapes}), we derive the \nhp 
relative abundance $X$(\nhp), using the \nhp and H$_2$ column 
densities obtained above, as:
\begin{eqnarray}
X(\mathrm{N}_2\mathrm{H}^+) \ = \
\frac{N\mathrm{(N}_2\mathrm{H}^+)}{N\mathrm{(H}_2)} \ = \ & 4.8\times10^{-11}.
\label{eq:abundace}
\end{eqnarray}

% The $X($\nhp) value obtained above is consistent, within the scatter, with 
% other estimates in massive Galactic star formation regions 
% in the ranges of \text{(1.6 - 3.8)\,$\times$\,10$^{-10}$}
% \citep[][Sandoval-Garrido et al. in prep.]{caselli2002,henshaw14}.

The $X($\nhp) value obtained above appears lower than
typical estimates in massive Galactic star forming regions
reported in different works 
\citep[\text{$[1.6 - 3.8]\,\times\,10^{-10}$};][Sandoval-Garrido 
et al. in prep.]{caselli2002,henshaw14}.
We consider it a good agreement considering the uncertainties of the involved
measurements (i.e. column density estimates).
For a comparison of the \nhp relative abundance between regions composing
the V-shapes and the rest of protocluster we would require to model the
full \nhp line emission in the whole region.

\begin{table}
\caption{\nhp full line fitting parameter ranges}
\label{table:n2hp_full}
\centering
\begin{tabular}{l l l l l}
\hline\hline
Excitation temperature [K]    & (T$_{\rm{ex}}$) &   2.73~$-$~80 & \\
Opacity                       & ($\tau$)        &   0.01~$-$~40 & \\
Centroid velocity     [km s$^{-1}$]  & ($v$)           &\, $-$25~$-$~15 & \\
Velocity dispersion   [km s$^{-1}$] & ($\Delta v$)    &   0.20~$-$~3  & \\
\hline
\end{tabular}
\tablefoot{Ranges used for the full \nhp line fitting.}
\end{table}

\section{Discussion}
\label{sect:discuss}
\subsection{V-shaped VGs in the literature}
The V-shaped velocity gradients described in this work have been detected
across multiple Galactic star forming system. 
\citet[][]{stutz16} introduced the Slingshot mechanism in the
Integral Shaped Filament (ISF) located in Orion A.
They show undulations of the region in both position and velocity, 
suggesting that these features appear to be ejecting protostars 
(see their Fig.~2). Furthermore, \citet[][]{stutz18} characterize a 
standing wave in the neighborhood of the ISF, consistent with the 
Slingshot mechanism. It is possible that the undulations in the 
works above might result in the observed V-shaped structures seen 
in different studies (see below).
\citet[][]{gonzalez19} identify six evenly spaced (every $\sim$\,0.44\,\pc)
velocity peaks along the spine of the ISF in Orion A. 
They suggest that this periodicity is consistent with the wave-like 
perturbation in the gas caused by the Slingshot mechanism. In \citet[][]{alvarez21} they analyze the L1482 
filament located in the California Molecular Cloud. 
In all of the analyzed tracers there is a clear velocity peak in their 
north region (length $\sim$~1.8\,pc, mass $\sim10^3$\,\msun). This subregion 
contains a higher gas density and higher number of YSOs compared to the more 
quiescent south part.

While the two regions described above are considered nearby 
(both at D\,$\lesssim$\,500\,pc), more distant regions also present 
these velocity features which we list below.
\citet[][]{atomsxi} study the velocity profiles along filaments from the 
ATOMS survey \citep[][]{atomsi}. The median mass of their sources is 
$\sim$~1.4\,$\times\,10^3$\,\msun with a median length of the filaments 
at $\sim\,1.35$\,pc. 
By analyzing the H$^{13}$CO$^+$~(1$-$0) emission they find converging 
VGs along filaments (see their Figs.~6 \& 10), which they also detected 
using simulations from \citet[][]{gomez14}. 
These VGs at scales comparables to the V-shapes presented here are consistent
with our VGs estimates (see their Fig.~7 \& 8). 
In \citet[][]{zhou23} they analyzed $^{13}$CO~(2$-$1) APEX/LAsMA data of the 
G333 complex. They identify multiple V-shaped VG (see their Fig.~7) which
they describe as the PV projection of a funneling structure in PPV space
(see their Fig.~9). The origin of this structure is due to material
inflowing towards the central hub and also due to gravitational contraction
of star-forming clouds or clumps. 

\citet[][]{redaelli22} use ALMA \nhp (1$-$0)
isolated component data of the high-mass (5200\,\msun)
clump AGAL014.492-00.139 identifying multiple coherent structures in 
PPV space (trees ``B'' and ``G''; right panel of their Fig.~7 \& 9).
These are characterized by multiple undulations, and possible V-shaped VGs. 
For their ``G'' PV distribution, they suggest that one scenario is where
the dense gas is flowing along the filament (of length $\sim$~0.2\,pc) 
from protostar ``p3'' towards the protostar ``p2''. This motion has
an \mrate\,=\,2.2\,$\times\,10^{-4}$~\msun~yr$^{-1}$, being in the range of the
\mrate we derive for our VGs (see Table~\ref{table:v-shapes}).

In \citet[][]{rawat24} they analyze $^{13}$CO(1$-$0) data 
obtained with the Purple Mountain Observatory, as part of the 
Milky Way Imaging Scroll Painting survey.
They detect a V-shaped VG (see their Fig.~14) along the ridge 
of the G148.24+00.41 (G148) cloud. 
This V-shape peaks towards the dense clump at the center of this 
region, possibly indicating gas inflow along their filaments F2 and F6 
towards the hub. Note that the length of the V-shape in G148
is $\sim15$\,pc, while our most prominent V-shape 
(Fig.~\ref{fig:main_gradient_blue}) is $\sim\,0.2\,\pc$.
Also the masses and lengths of their identified filaments are 
(1.3 to 6.9)$\,\times\,10^{3}$\,\msun and 14 to 38\,pc respectively, 
large compared to the total mass (2.5$\,\times\,10^3$\,\msun) and
extent of G353 ($\sim1.2$\,pc). This difference in probed
lengths and masses might be reflected by their mean VG~$\sim0.05$~\kmspc,
$\sim2.5$ orders of magnitude smaller than our VGs. 
This is consistent with the analysis presented in 
\citet[][]{atomsxi,zhou23}, where they observe an inverse relation 
between the spatial scale of a region and their velocity gradients 
(see their Fig.~8).

In \citet[][]{pan24} using APEX \co(2$-$1) data of the filamentary 
cloud G034.43+00.24 (G34) they identify converging VGs of 
lengths\,$\sim\,1$\,pc towards the ``middle ridge'' see their Fig.~3, 
top panel). They interpret these VG as gas flowing from the filaments 
onto dense clumps, located at the center of G34. 
These VGs of their southern and northern filaments are in the range of 
$\sim0.3-0.4$\,\kmspc, and they estimate the total mass inflow rate 
towards the middle ridge as $\sim5.5\,\times\,10^{-4}~$\msun\,~yr$^{-1}$, 
similar to our \mrate estimates.

Current work by Sandoval-Garrido et al. (in prep.) in G351.77 
\citep[intermediate protocluster, located at 2~kpc;][]{motte2022,reyes24}
use a similar analysis as we present in this work, where they
identify multiple V-shaped velocity structures.
In Salinas et al. (in prep) they analyze the kinematics of the evolved 
protocluster G012.80 \citep[located at 2.4~kpc;][]{motte2022}, where they 
implement similar techniques and find velocity signatures of filamentary 
rotation.

\subsection{Filamentary 3D morphology}

V-shaped VGs appear to be a generic feature across a wide
range of star forming environments, probing VGs with differences 
of up to $\sim\,2$ orders of magnitude in spatial scales ranging from
0.1 to $\sim\,$10\,pc. Despite being commonly detected in 
recent studies, it is still not clear how they are produced.
\citet[][]{henshaw14} highlights the degeneracy regarding the opposite 
interpretations of these V-shaped velocity structures.
They suggest that these VGs can be a signature of gas flows along 
kinked filaments towards a core located at their convergence point.
From our analysis regarding the most prominent V-shape 
(see \S~\ref{sect:vel_grad} \& \ref{sect:collapse}) we see that no core is 
located at its apex, although cores 2 \& 3 are within $\sim\,0.05$\,pc. 
Also the spatial offset between the center of the V-shape with the 
barycenter between these two cores is $\sim\,600$\,AU. 
This is consistent with the idea of small-scale gravitational collapse 
within the protocluster, similar to clump decoupling from their parent 
molecular cloud \citep[][]{peretto23}.
Based on this, we conclude that cores may be located in the vicinity of the 
velocity apex, and not necessarily on top of it. These gas flows towards
denser regions may result in the formation of high-mass cores in later
stages during the evolution of the protocluster.

Regarding the kinked morphology of the regions hosting V-shapes, 
one scenario regarding magnetized shocks is presented in 
\citet[][]{inoue2013} and \citet[][]{inoue2018}. They use 
magnetohydrodynamics simulations 
to characterize the interaction of molecular clouds and a 
magnetized shock produced by a cloud-cloud collision. They find that 
the shock layer decelerates as it collides with denser regions. 
This deceleration reshapes the shock layer to be oblique, 
leading to the formation of kinked filaments and converging flows, 
which are oriented towards the apex of these filaments. 
They predict that magnetic fields present in the region should be 
perpendicular to these filaments and bend with the shock around 
the filament \citep[][see their Fig.~3]{inoue2018}.
In \citet[][]{bonne2020} and \citet[][]{bonne2023} they propose that 
this scenario takes place in the Musca and the DR21 filaments. 
In both of these regions they detect V-shaped VGs which they suggest
are the result of cloud-cloud collisions bending the magnetic field 
\citep[][see their Fig.~22 \& 23]{bonne2020}. Further
observations of magnetic field polarization in the POS, along with 
information along the line of sight is required to evaluate 
these models. 

Another possibility is that these kinked structures could be caused by 
mechanisms such as the Slingshot. This mechanism proposes a standing 
wave, longitudinal gravitational instabilities, or large scale 
oscillations possibly caused by a possibly helical magnetic field morphology, 
causing ejections of protostars and protoclusters from their maternal filament 
\citep[][]{stutz16,stutz18,stutz18b,liu19}.

A different interpretation is that they are the product of out-flowing material 
coming from a forming protostar interacting with the surrounding dense gas 
(see their Fig.~12). 
To shed light into this degeneracy in G353 we compare the \nhp 
(dense gas tracer) radial velocities and SiO (shock/outflow tracer) 
as a proxy for energies.
The velocity range $\Delta$V covered by the \nhp emission is
$\sim$~8\,\kms, while for SiO is $\sim$~80\,\kms.
Given the difference in probed velocities between SiO and \nhp 
(see \S~\ref{sect:vel_grad}) and the analysis presented in 
\S~\ref{sect:collapse} we suggest that the V-shapes in G353 are a
signature of infall. 

The multiple VGs that conform the V-shapes present in G353 have 
values of \text{$\sim\,8$~to~$\sim\,31$\,\kmspc}, with timescales ranging 
from $\sim\,35~$to$~173$\,kyr, and \mrate values
\text{$(\sim\,0.4\,-\,9)\,\times\,10^{-4}~\msun~{\rm yr}^{-1}$}. These
values are similar to VGs in other regions with comparable sizes and masses. 
In G353 it is likely that these VGs are the result of dense 
gas moving through filaments, possibly increasing the density of the central 
regions, shaping the overall velocity field at large and small scales, and
leading to a further increase of the core population and star formation 
activity.

\subsection{Timescales and mass accretion rates}

One important aspect of the V-shapes that is still not 
well understood is the timescale associated of the VGs ($t_{VG}~=~$VG$^{-1}$).
It is not clear if nor how these timescales determine core formation 
lifetimes or impact the star formation environment in general.
In our sample of V-shapes the timescales are in tens of kyrs with the
average value of $\sim~67$\,kyr, $\sim~2$ times the cores $t_{ff}$, while the 
$t_{ff}$ of the whole protocluster is $\sim~0.21$\,Myr. 
In \citet[][]{rawat24} they estimate the longitudinal collapse 
timescales for their filaments, being in the range of 5\,$-$\,15\,Myr.
Using their derived VGs we estimate their associated timescales 
to be between $\sim16$~and~$\sim$\,50~Myr, $\sim1\,-\,2$ orders of magnitude 
larger than our small scale V-shapes timescales. We suggest that the 
VG timescales might serve as an upper limit for filamentary collapse timescales.
In \citet[][]{atomsxi} they determine gas accretion times as a function 
of the lengths of their filaments, assuming that the VGs produced by gas 
inflow (see their Fig.~11). At filament lengths comparable to our V-shapes 
($\sim~0.1$\,pc) their gas accretion timescales are on the order of 
our estimates (see Table~\ref{table:v-shapes}).  

\subsection{Depletion timescales}

It is also interesting to consider the mass accretion rates measured
here compared to the available protocluster mass reservoir to explore
implications for the duration of the gas dominated phase.  The total
mass accretion rate of our V-shaped structures is \text{$\dot{M}_{{\rm
in,\ Tot}}\,=\,$3$\,\times\,$10$^{-3}$\,\msun yr$^{-1}$} (see
Table~\ref{table:v-shapes}).  Considering the total mass (M$_{{\rm Tot}}$) of 
G353 as a mass reservoir, we estimate the 
depletion timescale ($t_{dep}$) as the time needed fully consume the gas. 
Here we assume that $\dot{M}_{{\rm in,\ Tot}}$ is
representative of flows feeding gas onto cores.  We estimate
$t_{dep}\,=\,$M$_{{\rm Tot}}/\dot{M}_{{\rm in,\ Tot}}$, where the total mass
of the region is 2500\,\msun \citep[][]{motte2022}. We obtain a
$t_{dep}$\,=\,0.8\,Myr, of similar magnitude but about four times
larger than the $t_{ff}$ of the protocluster. Considering that our
estimate of $\dot{M}_{{\rm in,\ Tot}}$ is certainly a lower limit (see
discussion above), the actual value value of $t_{dep}$ is likely to
be shorter, so closer to the free-fall time estimate. Given that the
protocluster does not appear to be in a state of free-fall (see
\S~\ref{sect:collapse}) but instead undergoing comparatively slow
gravitational contraction, the similarity in these relatively crude estimates 
seems remarkable. While we do not yet have an explanation for why 
relatively good match in timescales, it would seem to indicate that
protocluster evolution may be a self-regulating process.  Larger
samples and similar analysis will test this hypothesis.

Moreover, the approximate concordance of $t_{dep}$ and $t_{ff}$ may
indicate that the ``phase transition'' of protocluster gas mass being
converted into stellar mass could contribute a relevant ``negative
pressure'' counteracting effects of e.g.\ feedback over the lifetime
of the protocluster.

\section{Summary \& conclusions}
\label{sect:conclusions}
We characterize the complex dense gas kinematics of G353
using ALMA-IMF LP observations. The data used in this paper
mainly consist of the fully combined \nhp data cube, but we also include
1.3~mm continuum cores and DCN cores velocity catalogues,
SiO 12~m observation, and a $N$(H$_2$) 1.3~mm continuum
derived map. We summarize our main results below.
\begin{enumerate}
\item With our \nhp isolated component modeling, we find that
most of the 1.3~mm cores are located in regions with 2 to 3
velocity components. This indicates kinematic complexity down to
$\sim$4~kau scales.
\item We increase the number of cores with a \vlsr estimate in this region
by further examining the DCN emission and comparing it with the \nhp data
extracted towards the core positions.
We find that 11 cores, which were previously undetected in the
in the DCN background-subtracted fitting from \citet[][]{cunningham2023},
are identified with our method.
With this approach we increase our core velocities sample from 
15 to 26, accounting for $\sim\,58\%$ of the total 45 1.3~mm continuum cores.
These are presented in Table~\ref{table:core_vels}.
\item We show that the traditional PV diagram highlights large,
protocluster scale kinematics. In contrast, the intensity-weighted
PV diagram allows us access to the small, core scale dynamics
(see Figs.~\ref{fig:traditional_pv} \& \ref{fig:panel_pv}).
\item From the PV diagrams, we see the DCN core velocities are in 
agreement with the \nhp velocity distribution 
(within a few DCN channel widths). This suggests coupling 
between the cores and the dense gas in which they formed.
\item In the intensity-weighted PVs we see clear V-shaped 
velocity structures, composed by two linear velocity gradients (VGs)
converging into a common point. These VGs are present across 
all \nhp velocity components. Some of them are near the 
location of cores in both position and velocity 
(see \S~\ref{sect:pv_diagrams})
\item We successfully characterize nine V-shaped VGs well
detected in our \nhp data (see Figs.~\ref{fig:vshapes_pos}~\&~\ref{fig:v-shapes}). These structures are located mostly at the center of the
protocluster, where three filaments converge.
\item V-shape ``C'' (see Fig.~\ref{fig:main_gradient_blue}, 
\ref{fig:pv_angles}, and \ref{fig:multi_tracer}) is the most prominent 
across our sample. It is centered between cores 2 and 3, two of the 
most massive cores in this region. 
\item We estimate the barycenter of cores 2 \& 3, presenting an offset 
relative to the center of the V-shape of $\sim0.3$\arcsec ($\sim$\,600\,au)
well below the beam size of our \nhp data.
\item For V-shape ``B'' we find that core ``7'', with a mass
of 6\,\msun, is located within a $\sim$beam size from its apex.
\item We suggest that the dense gas is flowing along the filament, 
producing the V-shaped structure towards the derived barycenter.
\item  We characterize the VGs composing our sample of V-shapes by applying 
linear fits to these distributions. We estimate timescales associated to 
the VGs as $t_{VG}~=~$VG$^{-1}$. 
These timescales are between $\sim$35 to $\sim$170~kyr, with an average of 
$\sim$67~kyr. These values are short compared to the t$_{ff}$ of the 
protocluster ($\sim0.21$\,Myr), and $\sim2$ times larger that the 
cores average t$_{ff}$ ($\sim32$\,kyr).
\item We suggest that at small scales the \nhp V-shaped structures indicate 
gas motions along filaments, towards denser regions. Thus we interpret 
$t_{VG}$ as inflow timescales.
\item Using an H$_2$ mass map and the V-shapes mean timescales, 
we derive H$_2$ mass accretion rates of 
$(0.35\,{\rm to}\,8.77)\,\times\,10^{-4}~\msun~{\rm yr}^{-1}$, consistent 
with previous studies on regions that present gas flows along
filaments towards denser object or regions, such as protostars and clumps.
Moreover, V-shapes ``H'', ``C'', ``F'', and  ``B'' present
the largest \mrate(H$_2$) and they are located near or at the 
convergence point of the filaments (see Fig.~\ref{fig:vshapes_pos}).
\item In SiO, the PV structure covers a velocity range 
($\Delta$V) of $\sim~80$\,\kms, while for \nhp $\Delta$V is 
$\sim8$\,\kms. This difference suggests that \nhp is tracing infall, a
less energetic processes compared to SiO, a shock and outflow tracer.
\item We model the protocluster as a gravitationally collapsing 
sphere. The derived radial velocities are consistent with the large scale 
morphology of the traditional PV diagram. This agreement suggests that at 
large scales the G353 protocluster is undergoing gravitational contraction. 
\end{enumerate}

Overall, it is imperative to replicate the kinematic analysis presented
in this work in the remaining ALMA-IMF fields and other Galactic star
forming regions. By increasing the sample of analyzed fields we might
find correlations between evolutionary state \citep[young, intermediate, or
evolved; see][]{motte2022}, star formation activity, cores and
outflow population properties, and their velocity field. This approach
will allow us to better describe the kinematic processes taking place in 
this early stage of star formation.

\begin{acknowledgements}
The authors thank the referee for helpful comments that
improved the text.
This paper makes use of the following ALMA data: ADS/JAO.ALMA\#2017.1.01355.L. 
ALMA is a partnership of ESO (representing its member states), NSF (USA) and 
NINS (Japan), together with NRC (Canada), MOST and ASIAA (Taiwan), and KASI 
(Republic of Korea), in cooperation with the Republic of Chile. The Joint ALMA 
Observatory is operated by ESO, AUI/NRAO and NAOJ.

We thank Elena Redaelli, Diego R. Matus Carrillo, and Vineet Rawat
for very helpful discussions.

R.A. gratefully acknowledges support from ANID Beca Doctorado Nacional
21200897.

A.S. gratefully acknowledges support by the Fondecyt Regular 
(project code 1220610) and ANID BASAL project FB210003.

F.L. acknowledges the support of the Marie Curie Action of the European 
Union (project \textsl{MagiKStar}, Grant agreement number 841276)

F.M. acknowledges support from the French Agence Nationale de la 
Recherche (ANR) under reference ANR-20-CE31-009, of the Programme 
National de Physique Stellaire and Physique et Chimie du Milieu 
Interstellaire (PNPS and PCMI) of CNRS/INSU (with INC/INP/IN2P3).

R.G.M. and D.D.G. acknowledge support from UNAM-PAPIIT project 
IN108822 and from CONACyT Ciencia de Frontera project ID 86372. 

F.M., F.L., and N.C. acknowledge support from the European Research
Council (ERC) via the ERC Synergy Grant ECOGAL (grant 855130). 

N.C. acknowledges funding from the ERC under the European Union’s 
Horizon 2020 research.

P.S. was partially supported by a Grant-in-Aid for Scientific Research 
(KAKENHI Number JP22H01271 and JP23H01221) of JSPS. 

M.B. is a postdoctoral fellow in the University of Virginia's VICO 
collaboration and is funded by grants from the NASA Astrophysics Theory 
Program (grant number 80NSSC18K0558) and the NSF Astronomy \& 
Astrophysics program (grant number 2206516).

A.G. acknowledges support from the NSF under grants AAG 2008101 and CAREER 
2142300.

T.Cs. has received financial support from the French State in the framework 
of the IdEx Université de Bordeaux Investments for the future Program.

S.D.R. acknowledges the funding and support of ANID-Subdirección de
Capital Humano Magíster/Nacional/2021-22211000.
    
T.B. acknowledges the support from S. N. Bose National Centre for Basic
Sciences under the Department of Science and Technology, Govt. of India.

G.B. acknowledges financial support from the grants PID2020-117710GB-I00 and 
CEX2019-000918 funded by MCIN/AEI/10.13039/501100011033.

A.K. and L.B. gratefully acknowledge support from ANID BASAL project FB210003.
         
F.O. acknowledge the support of the Ministry of Science and Technology 
of Taiwan, projects No. 109-2112-M-007-008-, 110-2112-M-007-023-, and
110-2112-M-007-034-.
\end{acknowledgements}

\bibliographystyle{aa}
\bibliography{ref.bib}

\appendix

\section{Filamentary identification with FilFinder}
\label{app:filfinder}
Here we describe the procedure to identify the main filamentary
structures presented in \S~\ref{sect:iso_extraction} 
using \texttt{FilFinder} \citep[][]{filfinder}.

\setcounter{figure}{0}
\renewcommand{\thefigure}{A.\arabic{figure}}
\renewcommand{\theHfigure}{A.\arabic{figure}}

\begin{figure}
\centering
\includegraphics[width=\columnwidth]{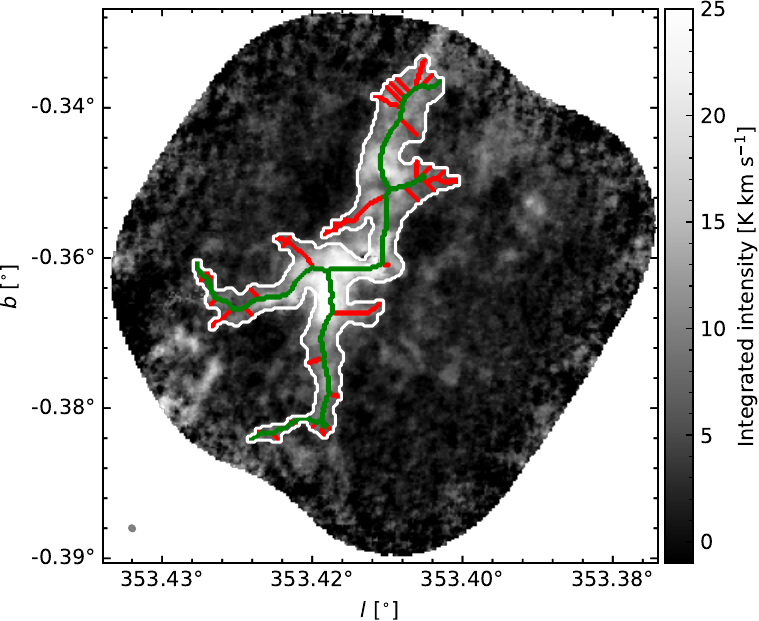}
   \caption{\texttt{FilFinder} filamentary identification. The
   background indicates the moment 0 map of the extracted \nhp 
   isolated components. The white contour shows the area where
   \texttt{FilFinder} identifies multiple filamentary structures
   (red and green lines). We remove the small scale structures (in red) 
   by ``pruning'' the skeleton structure from \texttt{medskel}, obtaining 
   the main filaments of G353. We represent these filaments with green lines.
   }
   \label{fig:filfinder}
\end{figure}

For this approach we use the moment 0 map of the extracted \nhp isolated
components that present a SNR\,$\geq$\,5. To estimate the moment 0 map
we used the \texttt{moment} task from the \texttt{SpectralCube} Python 
package, within the velocity range of $-31.5$\,\kms to 0\,\kms. 
As part of the pre-processing of the moment 0 map before the filamentary
detection, we decrease the contrast in the image by using the 
\texttt{preprocess\_image} task and its argument \texttt{flatten\_percent} 
set to 90. Now, in order to indicate to \texttt{FilFinder} the area 
we to identify filaments use the subtask \texttt{create\_mask} with 
the following parameters: 
\texttt{glob\_thresh: 4.5\,K\,km\,s$^{-1}$, size\_thresh: 0.25\,pc$^2$, 
smooth\_size: 0.12\,pc, border\_masking: False, 
fill\_hole\_size: 0.013\,pc$^2$}. The resulting mask is presented 
in Fig.~\ref{fig:filfinder} with a white contour.

Then, we obtain the skeletons of the mask by using \texttt{medskel}.
The derived structures are presented with red and green lines in 
Fig.~\ref{fig:filfinder}. Given we are only interested in the large scale 
filaments, we use \texttt{analyze\_skeletons} in order to ``prune'' the 
small scale structures.
For this pruning we use \texttt{branch\_thresh: 0.3\,pc,  
prune\_criteria: 'length', max\_prune\_iter: 0}. 
This approach results in removing the small filaments (red lines
in Fig.~\ref{fig:filfinder}) from the original skeleton and to obtain 
the main filamentary structure in G353 (green lines in 
Fig.~\ref{fig:filfinder}).

\section{Examples of the isolated components fitting}
\label{app:gaussian_fitting}
In \S~\ref{sect:iso_extraction} we decomposed the multiple isolated component
emission using \texttt{PySpecKit}. In Fig.~\ref{fig:iso_model} we present the
results of the Gaussian fitting for the high SNR spectra shown
in Fig.~\ref{fig:iso_extraction} (panels c, e, and f).

\section{DCN \& \nhp derived core velocity}
\label{app:nhp_core_vels}

In Table~\ref{table:core_vels} we provide the 1.3~mm core velocities
obtained from DCN \& \nhp data (see \S~\ref{sect:nhp_vels}), complementing
the published DCN core velocities catalogue from \citet[][]{cunningham2023}.
In the last column we indicate the number of \nhp velocity components 
detected in these cores.

\section{V-shaped structures}
\label{app:v-shapes}
\iffalse
{\bf In the left panel of Fig.~\ref{fig:vshapes_app_d} 
we show the spatial location of these V-shaped VGs. These are mostly 
located at the hub, where the three main filaments converge, and along 
the filament F3 (see \S~\ref{sect:fil_ident} and 
Appendix~\ref{app:filfinder}).
\fi

In \S~\ref{sect:vel_grad} we characterized the most prominent V-shaped
structure we detect in Fig.~\ref{fig:panel_pv}. We repeat this process for
other eight different V-shaped structures, including the linear fits to the
velocity gradients. 
In Fig.~\ref{fig:vshapes_app_d} we indicate the V-shapes 
location in PV space with dark points, arrows, and their IDs.
In Fig.~\ref{fig:v-shapes} we present individual close-ups for 
each V-shape. In Table~\ref{table:v-shapes} we list their VGs, timescales, 
H$_2$ masses, and mass accretion rates.

Here we list a few clarifications due to projection effects seen in 
these V-shaped structures:
\begin{itemize}
\item In position-position space, only V-shape ``B'' presents a core within 
a $\sim$beam size from its apex.
\item For V-shape ``A'', the 1.3~mm core with DCN single velocity
component, located  at the apex of this V-shape, is not spatially 
related to it.
\item In V-shapes ``G'' and ``H'' we see the same 1.3~mm cores with \nhp
velocities. These V-shapes are not the same distribution. They are
overlapped in PV space and spatially separated by $\sim10\arcsec$.
\item We improve the clarity of V-shape ``B'' by rotating the data in
PP space by 80\degree counter-clockwise. We apply this process for
V-shapes ``E'', ``G'', and ``H'' with an angle of 33\degree clockwise.
\item V-shapes G and H overlap in PV space but these are structures 
spatially separated.
\end{itemize}

\setcounter{figure}{0}
\renewcommand{\thefigure}{B.\arabic{figure}}
\renewcommand{\theHfigure}{B.\arabic{figure}}

\begin{figure*}
\centering
\includegraphics[width=\textwidth]{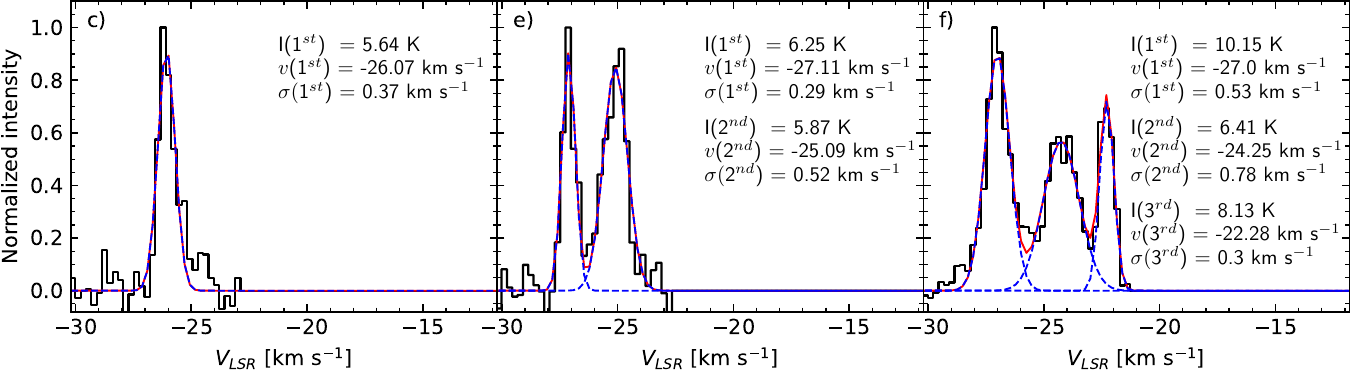}
   \caption{Gaussian velocity fits of the extracted \nhp isolated components.
   In black we show the high SNR isolated components from panels c), e), and 
   f) in Fig.~\ref{fig:iso_extraction}.
   The individual Gaussian components and the obtained model are 
   represented with dashed blue and solid red lines respectively. 
   On the right side of each panel we indicate the peak intensity (I), 
   the velocity centroid ($v$), and velocity dispersion ($\sigma$) of 
   each Gaussian component. 
   The notations 1$^{st}$, 2$^{nd}$, and 3$^{rd}$ indicate
   the Gaussian velocity components from left to right.}
   \label{fig:iso_model}
\end{figure*}

\section{SiO Intensity-weighted position-velocity diagram}
\label{app:sio_pv}

To create the SiO intensity-weighted PV diagram, first, we remove most of the
noisy spectra by considering data with $\rm{SNR}~\geq~2.5$.
Then, we estimate the integrated intensity and velocity centroid
at each pixel. We find improvements in our cleaning by using only spectra with
integrated intensity~$\geq~4$~K\,km\,s$^{-1}$.
Using the coordinate, integrated-intensity, and velocity centroid of each
spectrum, we create the SiO intensity-weighted PV diagrams we show in
\ref{fig:sio_panel_pv}.

\setcounter{figure}{0}
\renewcommand{\thefigure}{C.\arabic{figure}}
\renewcommand{\theHfigure}{C.\arabic{figure}}

\begin{figure}[h]
\centering
\includegraphics[width=\columnwidth]{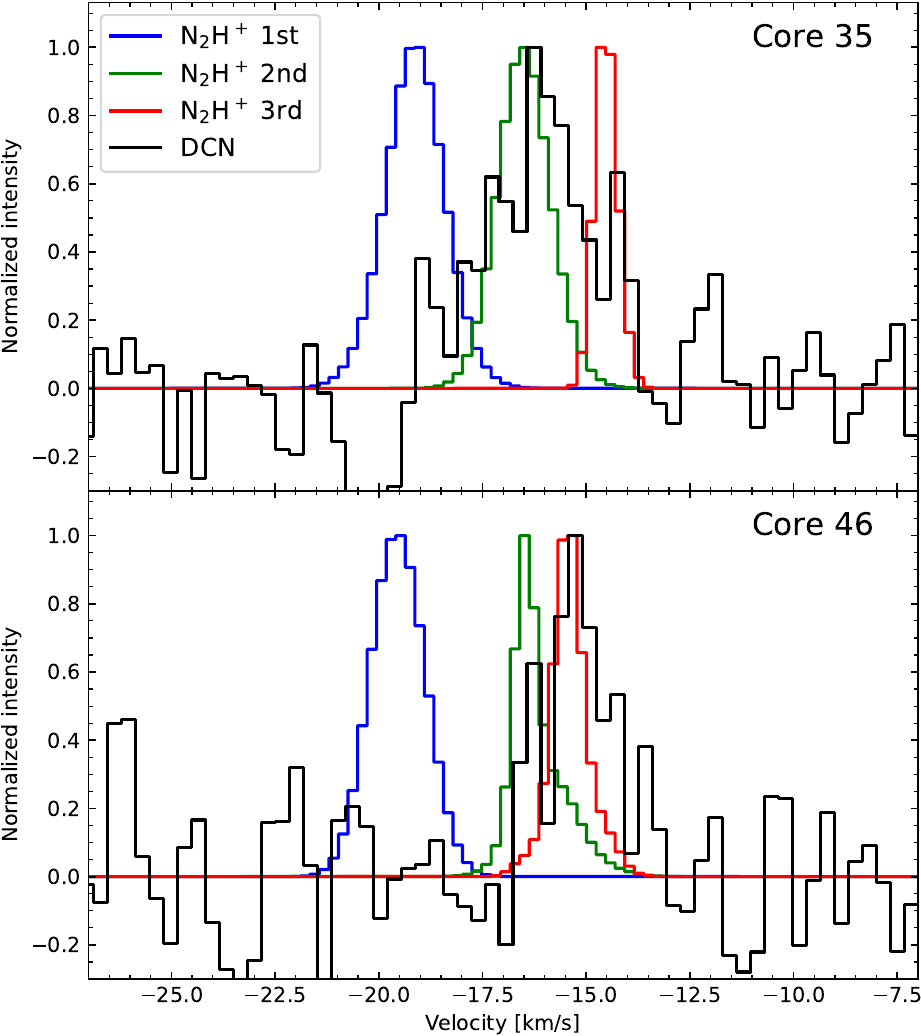}
   \caption{DCN and \nhp normalized mean spectrum of cores 35 (top)
   and 46 (bottom). We show the multiple \nhp isolated velocity components
   with blue, red, and green colors. We present the DCN emission in black.
   We see a match between the DCN emission and one of the \nhp
   velocity components. We determine the \nhp velocity for 11 cores
   with no DCN velocity fits. These are listed in Table~\ref{table:core_vels}.
   }
   \label{fig:n2hp_cores_vel}
\end{figure}

\setcounter{table}{0}
\renewcommand{\thetable}{C.\arabic{table}}
\begin{table*}[h]
\caption{1.3~mm core catalogue of DCN \& \nhp velocities}
\label{table:core_vels}
\centering
\begin{tabular}{rllNNNNNcclN}
\hline\hline
Core & \col{RA} & \col{DEC} & \col{\ \ \ F$_{\rm{A}}$} & \col{\ \ \ F$_{\rm{B}}$}& \col{PA} & \col{Mass} & \col{V$_{LSR}$} \hspace{-1cm} & & & \ \ \ \ {Type} & \col{Number of}\\                                                                                                  
Number & \col{[\degree]} & \col{[\degree]} & \ \ {[\arcsec]}    & \ \ {[\arcsec]} &  {[\degree]}  & {[\msun]} &  \col{[km s$^{-1}$]}  \hspace{-1cm} &   &   &  & \col{\nhp components} \\
\hline
2  & 262.6165032 & -34.6955865 & 1.98 & 1.59 & 64.00  & 20.7  & -20.45$\pm$0.065 & & & DCN, Single   &  2       \\
3  & 262.6184156 & -34.6965240 & 2.59 & 1.79 &146.00 &  9.4   & -16.48$\pm$0.040  & & & DCN, Single  &  2       \\
4  & 262.6103159 & -34.6932659 & 1.56 & 1.46 &104.00 &  5.2   & -16.53$\pm$0.155 & & & DCN, Single   &  3       \\
5  & 262.6101515 & -34.6960014 & 2.03 & 1.75 & 79.00  & 16.0  & -16.94$\pm$0.087 & & & DCN, Single   &  3       \\
6  & 262.6049155 & -34.6934384 & 1.60 & 1.48 &129.00 &  4.9   & -18.68$\pm$0.126 & & & DCN, Single   &  2       \\
7  & 262.6137738 & -34.6947298 & 1.63 & 1.27 & 80.99  &  6.0    & -19.79           & & & DCN \& \nhp &  3       \\
8  & 262.6039531 & -34.6936374 & 2.26 & 1.56 & 97.32  &  6.6 &       \novel     & & & \notype        &  \notypecom \\
9  & 262.6192359 & -34.6903650 & 2.69 & 1.96 & 62.96  &  3.7 & -16.09           & & & DCN \& \nhp    &  3       \\
11 & 262.6243189 & -34.6880780 & 2.12 & 1.95 &172.70  &  2.8 & -16.09           & & & DCN \& \nhp    &  3       \\
12 & 262.6072148 & -34.6969795 & 2.98 & 2.02 & 86.00  & 10.3 & -17.83$\pm$0.019 & & & DCN, Single    &  3       \\
13 & 262.6078228 & -34.6996836 & 1.91 & 1.48 &153.90 &  2.7  &       \novel     & & & \notype        &  \notypecom \\
14 & 262.6147937 & -34.6946762 & 2.00 & 1.58 &124.00 &  6.2 &  -14.36$\pm$0.090  & & & DCN, Single   &  3       \\
15 & 262.6107433 & -34.6964412 & 1.96 & 1.59 & 93.00  &  6.6 & -14.78$\pm$0.087 & & & DCN, Complex   &  \notypecom \\
16 & 262.6215941 & -34.6989408 & 2.68 & 2.49 & 19.37  &  1.5 &       \novel     & & & \notype        &  \notypecom \\
17 & 262.5954514 & -34.6916168 & 1.57 & 1.42 & 87.50  &  0.9 &       \novel     & & & \notype        &  \notypecom \\
18 & 262.5927434 & -34.7052494 & 1.97 & 1.56 & 89.26  &  0.8 &       \novel     & & & \notype        &  \notypecom \\
19 & 262.6064012 & -34.7019756 & 1.55 & 1.20 & 57.74  &  0.5 &       \novel     & & & \notype        &  \notypecom \\
20 & 262.6111096 & -34.6932787 & 1.67 & 1.63 & 178.00 &  1.3  &       \novel     & & & \notype       &  \notypecom \\ 
21 & 262.6131910 & -34.6939495 & 2.10 & 1.48 & 108.00 &  2.4 & -18.56$\pm$0.063 & & & DCN, Single    &  3       \\
22 & 262.6118441 & -34.6946150 & 1.89 & 1.59 & 96.65  &  1.7 & -16.76           & & & DCN \& \nhp    &  3       \\
23 & 262.6028175 & -34.6925438 & 1.84 & 1.31 & 112.70 &  0.8 &       \novel     & & & \notype        &  \notypecom \\
24 & 262.6198349 & -34.6960383 & 1.88 & 1.72 & 76.00  &  0.8  & -19.50$\pm$0.089 & & & DCN, Single   &  2       \\
25 & 262.6155222 & -34.6952591 & 1.87 & 1.20 & 137.30 &  2.5 &       -18.78     & & & DCN \& \nhp    &  2       \\
26 & 262.6143434 & -34.6917027 & 1.49 & 1.24 & 137.30 &  0.8 &       \novel     & & & \notype        &  \notypecom \\
27 & 262.6000802 & -34.6910324 & 3.39 & 2.51 & 48.16  &  1.8 &       \novel     & & & \notype        &  \notypecom \\
28 & 262.6253977 & -34.6999713 & 2.56 & 1.87 & 39.31  &  0.8 &       \novel     & & & \notype        &  \notypecom \\
29 & 262.6133074 & -34.6919187 & 1.67 & 1.26 & 76.67  &  0.7 &       \novel     & & & \notype        &  \notypecom \\
30 & 262.6114686 & -34.6962602 & 1.52 & 1.41 & 30.00  &  2.3 & -12.81$\pm$0.074 & & & DCN, Single    &  2       \\
31 & 262.6096651 & -34.6925680 & 1.72 & 1.34 & 119.10 &  0.7  &       \novel     & & & \notype       &  \notypecom \\
32 & 262.6094126 & -34.6910985 & 2.44 & 2.02 & 59.22  &  2.0 &       \novel     & & & \notype        &  \notypecom \\
33 & 262.6287106 & -34.6862068 & 2.32 & 2.00 & 16.64  &  1.7  &       \novel     & & & \notype       &  \notypecom \\
34 & 262.5982914 & -34.6919006 & 1.81 & 1.36 & 139.00 &  0.8 & -18.77$\pm$0.080  & & & DCN, Single   &  2       \\
35 & 262.6142011 & -34.6940134 & 2.31 & 1.81 & 111.00 &  3.4 & -16.09           & & & DCN \& \nhp    &  3       \\
36 & 262.6202758 & -34.7001995 & 2.19 & 1.88 & 141.00 &  0.8 &       \novel     & & & \notype        &  \notypecom \\
37 & 262.6010398 & -34.6950114 & 1.74 & 1.33 & 75.00  &  0.8 & -17.83$\pm$0.053 & & & DCN, Single    &  3       \\
38 & 262.5971034 & -34.6920396 & 2.01 & 1.61 & 104.80 &  0.7 &       \novel     & & & \notype        &  \notypecom \\
39 & 262.6054437 & -34.6963773 & 2.05 & 1.88 & 158.50 &  1.4 & -16.43           & & & DCN \& \nhp    &  3       \\
40 & 262.5917454 & -34.6897316 & 1.85 & 1.50 & 92.93  &  0.6 &       \novel     & & & \notype        &  \notypecom \\
41 & 262.6095777 & -34.6983259 & 2.23 & 1.83 & 72.51  &  1.3 &       \novel     & & & \notype        &  \notypecom \\
42 & 262.5975992 & -34.6876666 & 1.51 & 1.22 & 24.46  &  0.3 &       \novel     & & & \notype        &  \notypecom \\
43 & 262.6035166 & -34.6966807 & 2.48 & 1.78 & 95.00  &  1.2 & -17.33$\pm$0.062 & & & DCN, Single    &  3       \\
44 & 262.6030115 & -34.6956424 & 1.76 & 1.55 & 96.00  &  0.5 & -16.99$\pm$0.091 & & & DCN, Single    &  3       \\
45 & 262.6143008 & -34.6909376 & 3.40 & 2.75 & 94.93  &  2.6 &       \novel     & & & \notype        &  \notypecom \\
46 & 262.6187648 & -34.6912377 & 3.10 & 2.15 & 40.67  &  0.9 & -15.08 & & & DCN \& \nhp              &  3       \\              
47 & 262.6178453 & -34.6919943 & 3.09 & 2.48 & 45.43  &  1.6 &       \novel     & & & \notype        &  \notypecom \\
\hline
\end{tabular}
\tablefoot{Velocities of the 1.3~mm continuum derived cores.
The column ``Type'' indicates if the core velocity is determined by a
single or complex DCN spectra \citep[][]{cunningham2023}, or
using both DCN \& \nhp data (this work). In the last column indicate 
the number of \nhp velocity components inside these cores. 
For completeness we include the properties of the 1.3~mm cores 
with no velocity determinations. These cores present a ``---'' mark
in the last three columns. The core catalogue from Louvet et al.
(submitted) does not consider core ``1'' nor core ``10'' given
their selection criteria reject cores with free-free 
contamination or cores undetected at 3~mm.}
\end{table*}
% \end{document}

\setcounter{figure}{0}
\renewcommand{\thefigure}{D.\arabic{figure}}
\renewcommand{\theHfigure}{D.\arabic{figure}}

\begin{figure}[h]
\centering
\includegraphics[width=\columnwidth]{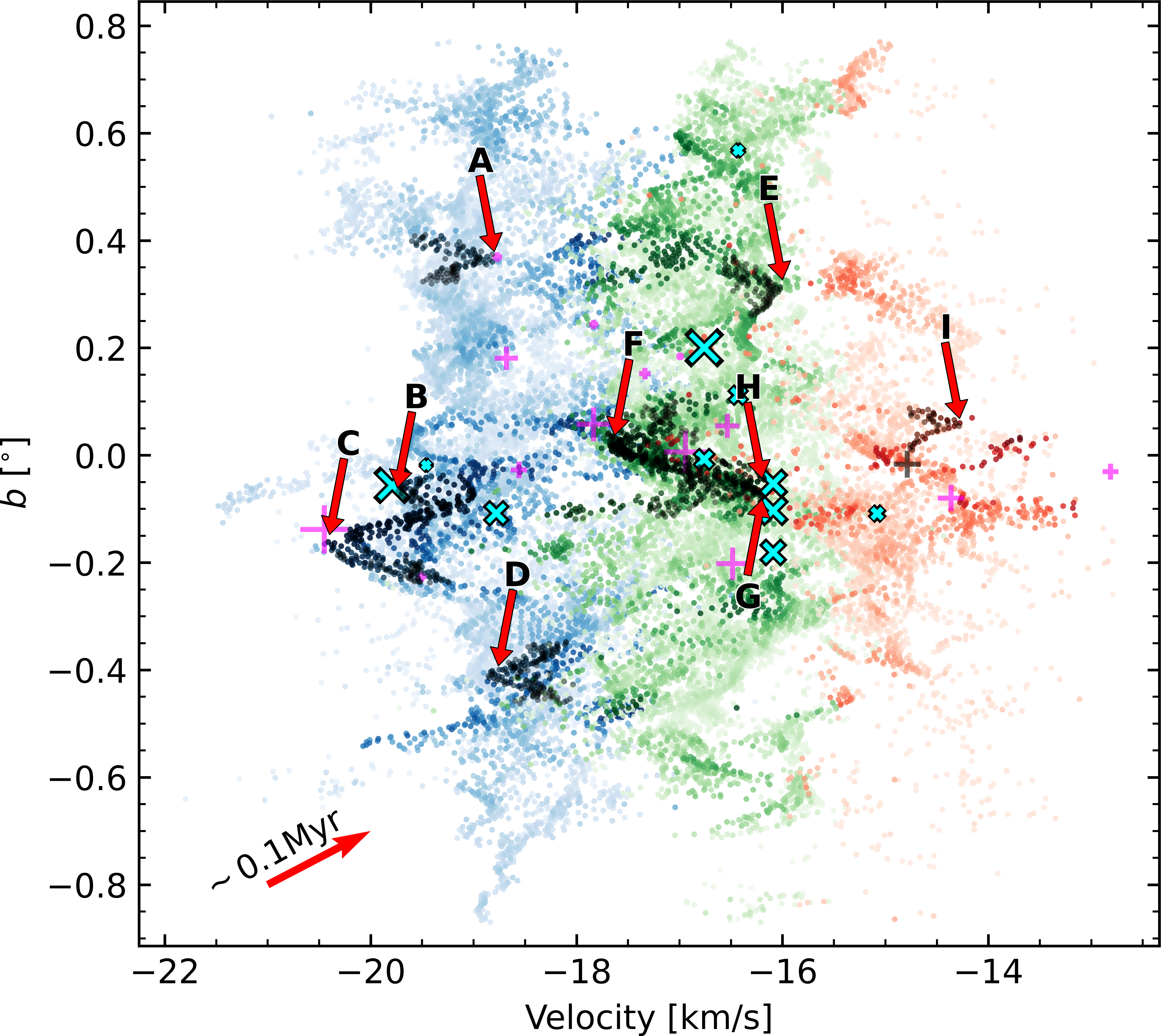}
\caption{
   V-shapes location in PV space. We highlight the V-shapes listed in 
   Table~\ref{table:v-shapes} with black points and indicate them
   with red arrows and their ID.
   The core velocities and the \nhp velocity distributions follow the
   same definitions from the top right panel in Fig.~\ref{fig:panel_pv}.
   V-shapes G and H overlap in PV space but these structures are spatially
   separated (left panel).
   }
\label{fig:vshapes_app_d}
\end{figure}

\begin{figure*}[ht]
\centering
\setlength{\tabcolsep}{1pt}
\begin{tabular}{c c}
\begin{overpic}[width = 0.44\textwidth]{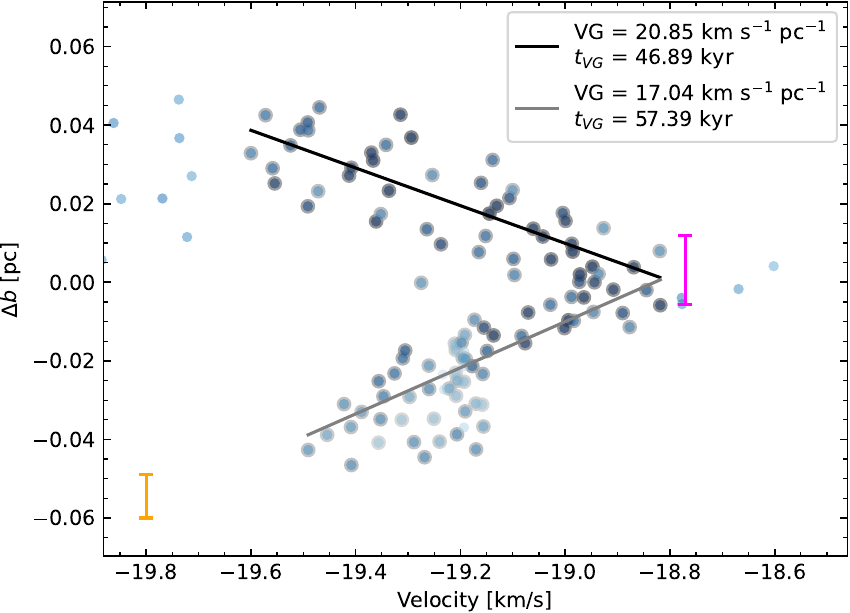}\put(15,66){\large A} \end{overpic} & 
\begin{overpic}[width = 0.44\textwidth]{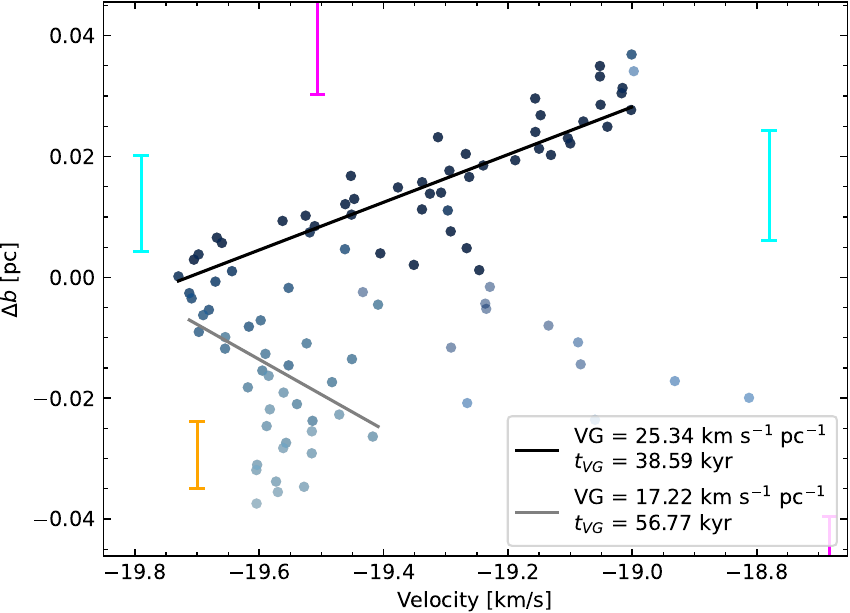}\put(15,66){\large B} \end{overpic} \\
\begin{overpic}[width = 0.44\textwidth]{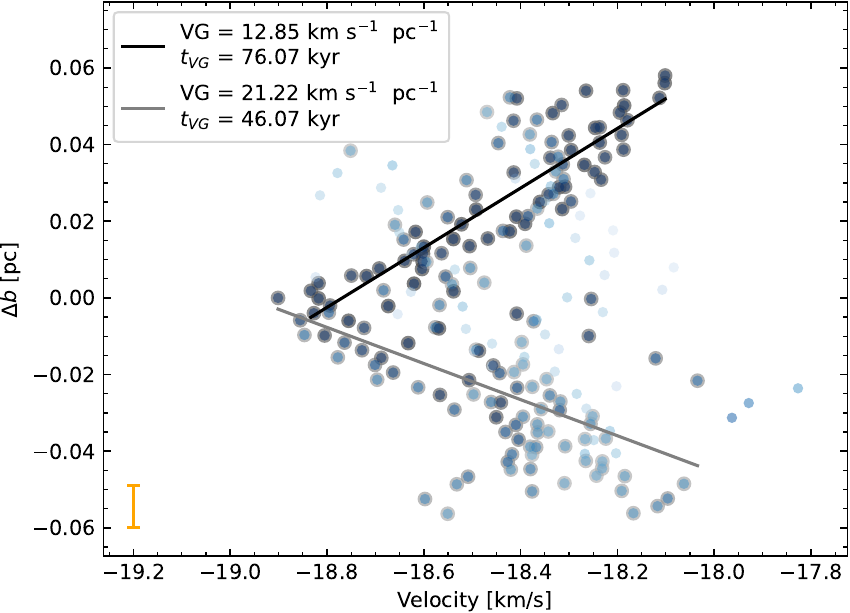}\put(17,10){\large D} \end{overpic} &
\begin{overpic}[width = 0.44\textwidth]{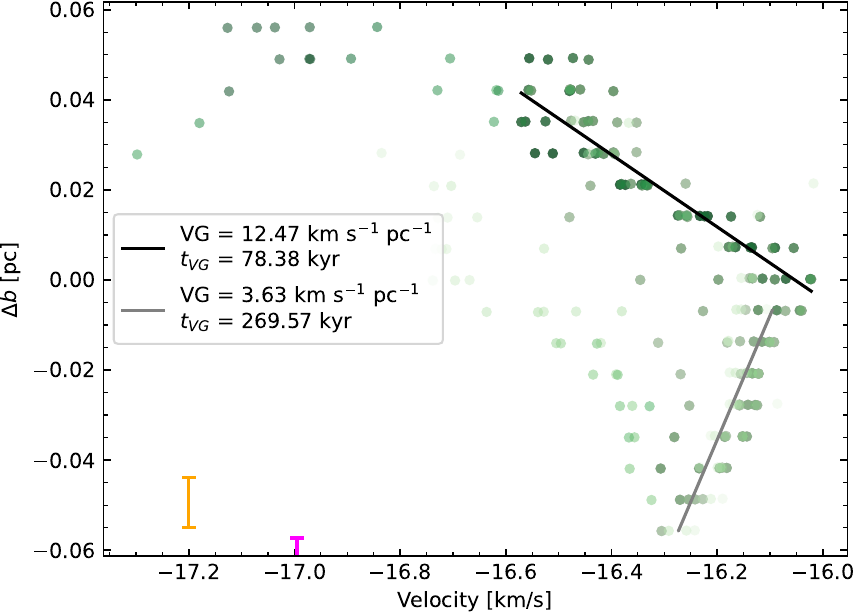}\put(15,10){\large E} \end{overpic} \\
\begin{overpic}[width = 0.44\textwidth]{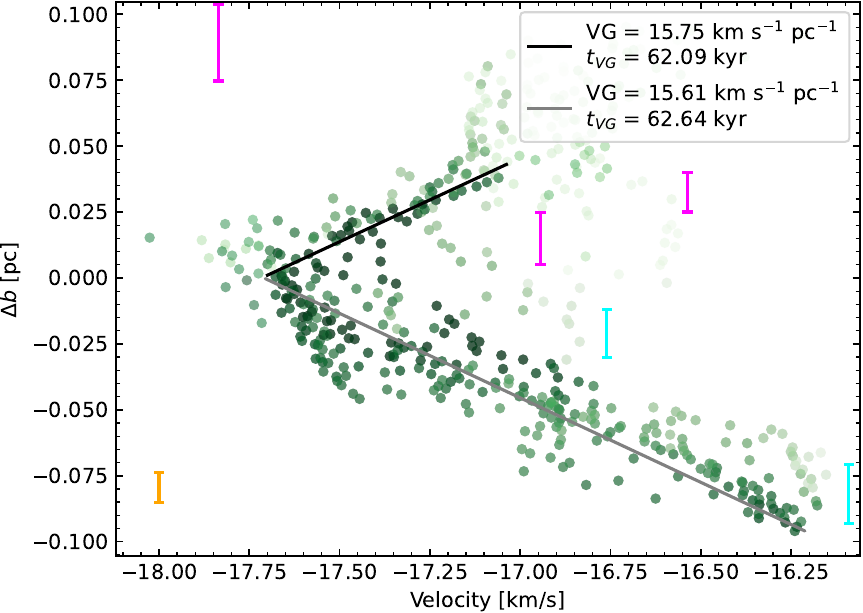}\put(15,10){\large F} \end{overpic} &
\begin{overpic}[width = 0.44\textwidth]{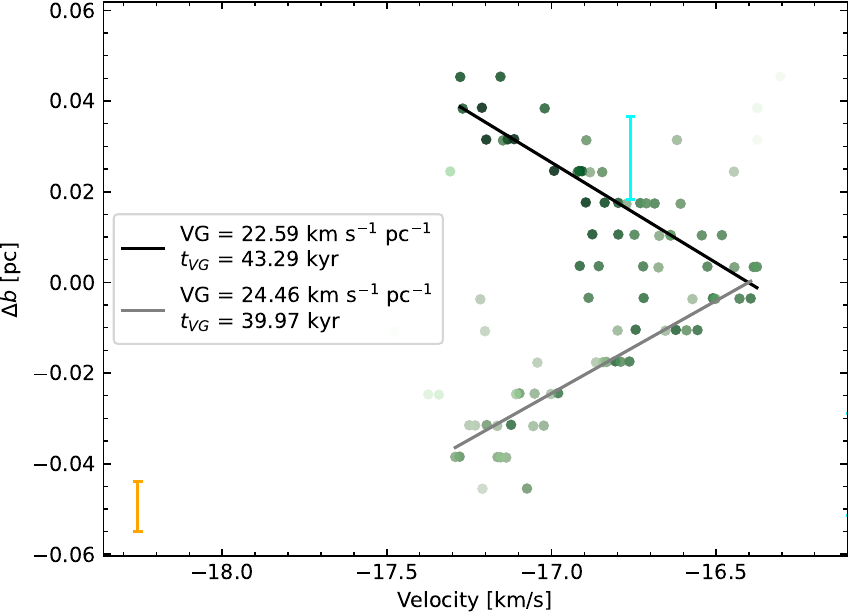}\put(17,10){\large G} \end{overpic} \\
\begin{overpic}[width = 0.44\textwidth]{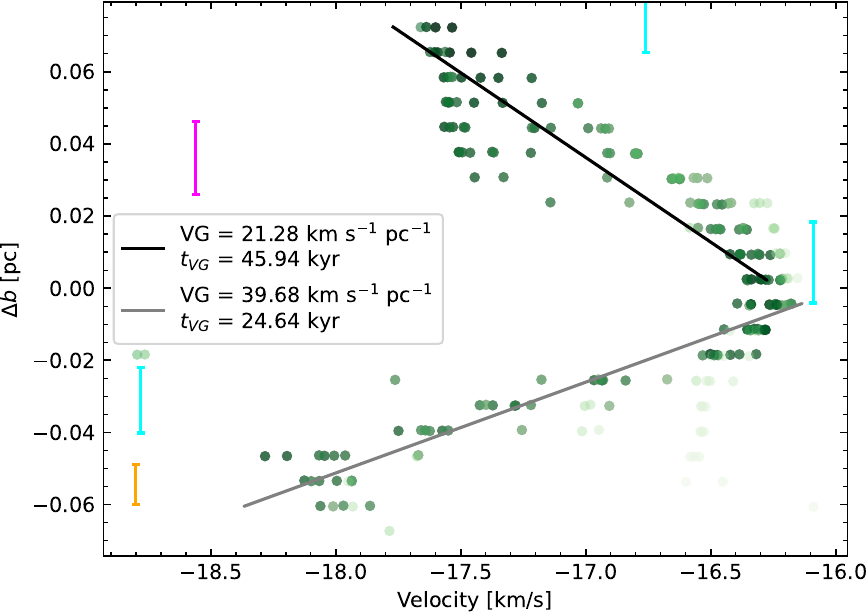}\put(15,66){\large H} \end{overpic} &
\begin{overpic}[width = 0.44\textwidth]{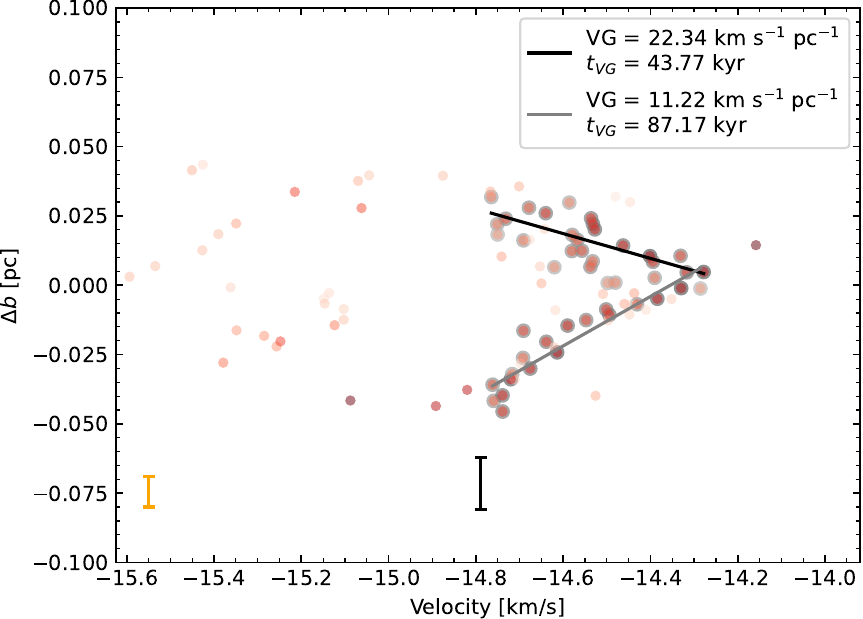}\put(15,66){\large I} \end{overpic} \\
\end{tabular}
\caption{V-shaped structures listed in Table~\ref{table:v-shapes}, with the
exception of ``C'' shown in Fig.~\ref{fig:main_gradient_blue}.
We indicate the ``V-shape ID'' from Table~\ref{table:v-shapes} at the
top/bottom left corner of each plot.
The colors of the distributions, DCN and DCN \& \nhp derived core 
velocities, and beam size follow the same color and marker convention 
from Fig.~\ref{fig:panel_pv}.
See Appendix~\ref{app:v-shapes} for clarifications regarding
projection effects on these diagrams.
}
\label{fig:v-shapes}
\end{figure*}

\setcounter{table}{0}
\renewcommand{\thetable}{D.\arabic{table}}

\begin{table*}
\caption{Characterized V-shaped structures}
\label{table:v-shapes}
\centering
\begin{tabular}{c c c N c N c N N }
\hline\hline
\\[-3mm]
V-shape ID & $l$ &  $b$ & {$M$(H$_2$)}  & Upper / lower VG & \col{Mean VG} & {Upper / lower} $t_{VG}$ & {$t_{VG \ mean}$}   &  {$\dot{M}_{\rm in}$(H$_2$)} \\
&  {[\degree]} & {[\degree]} & [M$_{\odot}$]& [km s$^{-1}$ pc$^{-1}$]    & {[km s$^{-1}$ pc$^{-1}$]}      & [kyr] &   [kyr]    & {[10$^{-4}$ \msun yr$^{-1}$]}\\% & \ [K]
\hline
A & 353.3981&-0.3506&  8.13 &  20.85\ /\ 17.04   &  18.95   & 46.89 \ /\ \ 57.39  &   52.14 & 1.56\\
B & 353.4127&-0.3632& 13.29&  25.34\ /\ 17.22   &  21.28   & 38.59 \ /\ \ 56.77  &   47.68 & 2.79\\
C & 353.4135&-0.3657& 53.02&  17.69\ /\ 13.26   &  15.48   & 55.28 \ /\ \ 73.75  &   64.52 & 8.22 \ \\
D & 353.4133&-0.3727& 7.96&  12.85\ /\ 21.22   &  17.15   & 76.07 \ /\ \ 46.07  &   61.07  & 1.31\\
E & 353.4096&-0.3521&  6.27&  12.47\ /\  \, 3.63 &   8.05   & \ 78.38 /\  269.57\, & 173.98& 0.36\\
F & 353.4128&-0.3604& 27.24&  15.75\ /\ 15.61   &  15.68   & 62.09 \ /\ \ 62.64  &   62.37 & 4.37\\
G & 353.4110&-0.3630&  3.24&  22.59\ /\ 24.46   &  23.53   & 43.29 \ /\ \ 39.97  &   41.63 & 0.79\\
H & 353.4140&-0.3627&  31.46&  21.28\ /\ 39.68   &  30.48   & 45.94 \ /\ \ 24.64  &   35.29& 8.91\ \ \ \\
I & 353.4091&-0.3595& 16.13&  22.34\ /\ 11.22   &  16.78   & 43.77 \ /\ \ 87.17  &   65.47 & 2.46 \\
\hline
\end{tabular}
\tablefoot{Properties of the nine, well characterized, V-shaped structures identified in our
\nhp data. The coordinates indicate the position of the velocity apex of 
each V-shape.
V-shape ``C'' represents the structure analyzed in \S~\ref{sect:vel_grad}.
We show the PV distribution of these structures in Fig.~\ref{fig:v-shapes}.}
\end{table*}

\setcounter{figure}{0}
\renewcommand{\thefigure}{E.\arabic{figure}}
\renewcommand{\theHfigure}{E.\arabic{figure}}

\begin{figure*}[b]
\centering
\includegraphics[width=\textwidth]{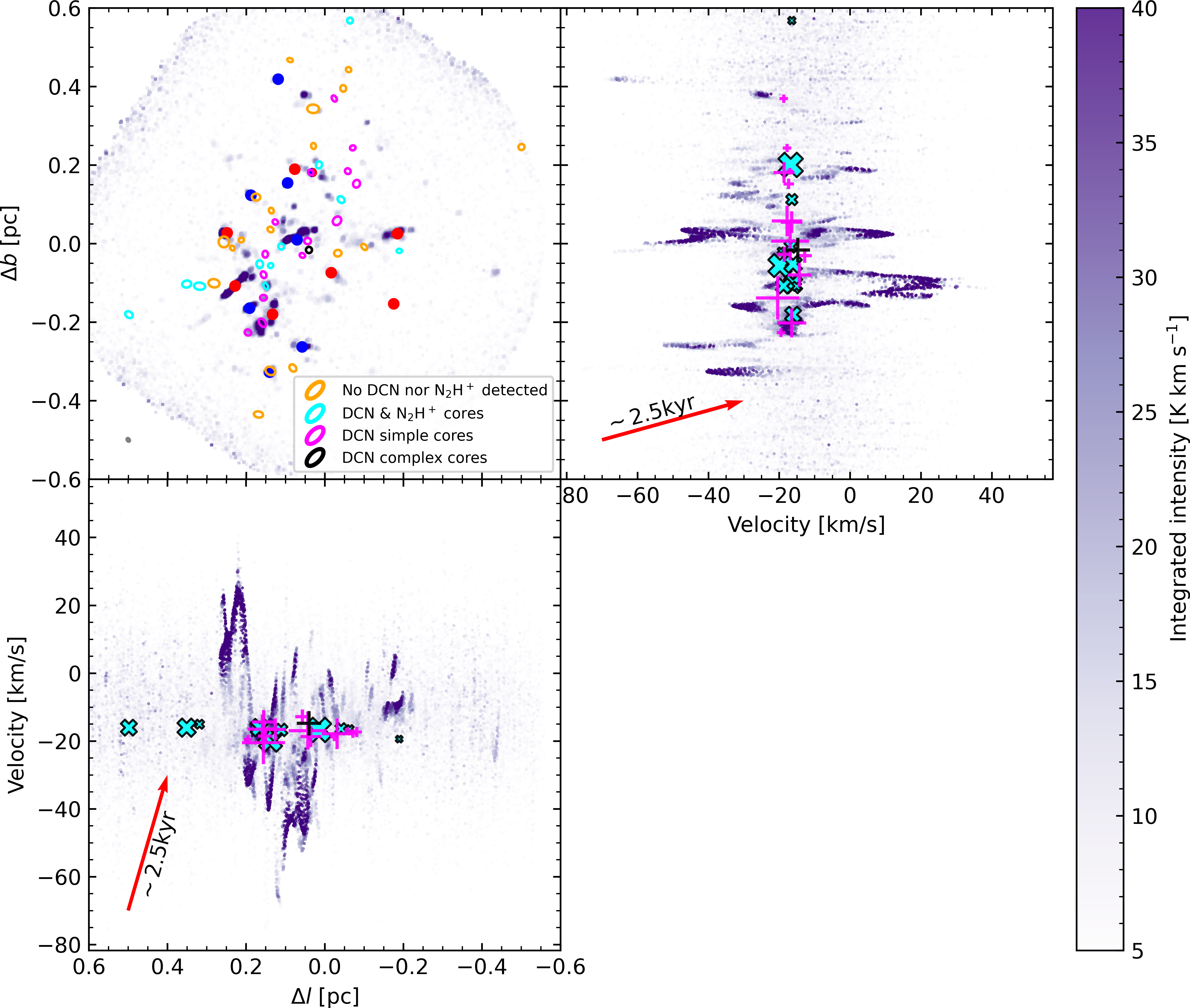}
   \caption{ALMA-IMF 12~m SiO equivalent of Fig.~\ref{fig:panel_pv} 
   using data from \citet[][]{cunningham2023}. For the cores, we use 
   the same marker and color convention from 
   Fig.~\ref{fig:panel_pv}. With filled blue, red, and `red+blue' 
   circles we represent the SiO outflow candidates \citep[][]{towner24}.
   With red arrows we indicate a VG~=~400\,\kmspc corresponding to
   a timescale $t_{VG}~=~2.5$~kyr. The velocity range ($\Delta$V)
   covered by the SiO emission is $\sim$80\kms, about 10 times the
   velocity range traced by \nhp. This velocity difference
   suggests that SiO is tracing processes (outflows) $\sim100$ times 
   more energetic ($e_k~=~\Delta \rm{V}/2$) than \nhp (possibly infall).
   }
   \label{fig:sio_panel_pv}
\end{figure*}

\section{G353 power law density profile }
\label{app:density_profile}

\setcounter{equation}{0}
\renewcommand{\theequation}{F.\arabic{equation}}

Here we provide the derivation of the density profile used
for the gravitationally collapsing sphere.
We assume a power law density profile defined as:
\begin{eqnarray}
\rho(r) & = & \rho_0 \left (\frac{r}{\rm{pc}} \right )^{-\gamma},
\end{eqnarray}
where $\gamma=5.65$ provides a good fit to the edges of the PV
distribution seen in Fig.~\ref{fig:traditional_pv}. 

To determine the value of $\rho_0$, we integrate this expression
in a sphere (Eq.~\ref{eq:int_den}), with $r_{min}~<~r<~0.5$~\pc.
Based on different tests, probing total masses from 
$50~-~10^3$\,\msun and $\gamma$ values from $2~-~6$, we set the 
total mass of the sphere to 150\,\msun.  
We define $r_{min}~\sim~0.007~{\rm pc}$ which corresponds to the pixel
size of the \nhp data at a distance of 2~kpc.
\begin{eqnarray}
M_{enc}(r~=~0.5~{\rm pc})&=&4\pi \rho_0 \int_{r_{min}}^{{\rm 0.5\,pc}}
\left (\frac{r}{{\rm pc}}\right)^{-\gamma} r^2dr \label{eq:int_den}\\
&=&4\pi \rho_0  \frac{r^{3-\gamma}}{3-\gamma}
\Bigg \rvert_{r~=~r_{min}}^{r~=~0.5~{\rm pc}}~{\rm pc}^{\gamma}\\
&=&\frac{4\pi \rho_0}{3-\gamma} (0.5^{3-\gamma}-{r_{min}}^{3-\gamma})~{\rm pc}^{\gamma},
\label{eq:den_prof_general}
\end{eqnarray}
where $M_{enc}(r\,=\,0.5\,{\rm pc}$)\,=\,150~\msun,
$r_{min}\,=\,7\times10^{-3}$\,pc, and $\gamma$\,=\,5.65,
from Eq.~\ref{eq:den_prof_general}, we obtain:
\begin{eqnarray}
\rightarrow \rho_0 & = & 6.1~\times~10^{-5} \frac{{\rm M}_{\odot}}{{\rm pc}^3}.
\label{eq:final_derivation}
\end{eqnarray}


\end{document}